\documentclass[fleqn,usenatbib]{mnras}

\usepackage{newtxtext,newtxmath}

\usepackage[T1]{fontenc}
\usepackage{gensymb}
\DeclareRobustCommand{\VAN}[3]{#2}
\let\VANthebibliography\thebibliography
\def\thebibliography{\DeclareRobustCommand{\VAN}[3]{##3}\VANthebibliography}

\usepackage{graphicx}	
\usepackage{amsmath}	
\usepackage{xcolor}

\title[gr8stars I]{gr8stars I: A homogeneous spectroscopic study of bright FGKM dwarfs and a public library of their high-resolution spectra}

\author[Freckelton et al.]{
Alix Violet Freckelton $^{1}$$^{\href{https://orcid.org/0009-0007-1053-0004}{\includegraphics[scale=0.5]{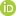}}}$\thanks{Email : axf859@student.bham.ac.uk},
Annelies Mortier $^{1}$$^{\href{https://orcid.org/0000-0001-7254-4363}{\includegraphics[scale=0.5]{orcid.jpg}}}$,
Megan Bedell $^{2}$ $^{\href{https://orcid.org/0000-0001-9907-7742}{\includegraphics[scale=0.5]{orcid.jpg}}}$, 
Sam Morrell $^{3}$ $^{\href{https://orcid.org/0000-0001-6352-5312}{\includegraphics[scale=0.5]{orcid.jpg}}}$,
\newauthor 
Tim Naylor $^{3}$ $^{\href{https://orcid.org/0000-0002-0506-8501}{\includegraphics[scale=0.5]{orcid.jpg}}}$,
Lars A. Buchhave $^{4}$ $^{\href{https://orcid.org/0000-0003-1605-5666}{\includegraphics[scale=0.5]{orcid.jpg}}}$,
Guy R. Davies $^{1}$ $^{\href{https://orcid.org/0000-0002-4290-7351}{\includegraphics[scale=0.5]{orcid.jpg}}}$,
J. I. Gonz\'alez Hern\'andez $^{5,6}$ $^{\href{https://orcid.org/0000-0002-0264-7356}{\includegraphics[scale=0.5]{orcid.jpg}}}$,
\newauthor
Baptiste Klein $^{7}$ $^{\href{https://orcid.org/0000-0003-0637-5236}{\includegraphics[scale=0.5]{orcid.jpg}}}$,
Ernst J.W. de Mooij $^{8}$ $^{\href{https://orcid.org/0000-0001-6391-9266}{\includegraphics[scale=0.5]{orcid.jpg}}}$,
Vera Maria Passegger $^{9,10}$ $^{\href{https://orcid.org/0000-0002-8569-7243}{\includegraphics[scale=0.5]{orcid.jpg}}}$,
Andreas Quirrenbach $^{11}$,
\newauthor
Arpita Roy $^{12}$ $^{\href{https://orcid.org/0000-0001-8127-5775}{\includegraphics[scale=0.5]{orcid.jpg}}}$,
Nuno C. Santos $^{13,14}$,
S\'ergio G. Sousa $^{13}$ $^{\href{https://orcid.org/0000-0001-9047-2965}{\includegraphics[scale=0.5]{orcid.jpg}}}$,
A. Su\'arez Mascare\~no $^{5,6}$ $^{\href{https://orcid.org/0000-0002-3814-5323}{\includegraphics[scale=0.5]{orcid.jpg}}}$,
\newauthor
Maria Tsantaki $^{15}$ $^{\href{https://orcid.org/0000-0002-0552-2313}{\includegraphics[scale=0.5]{orcid.jpg}}}$,
Lily L. Zhao $^{16}$ $^{\href{https://orcid.org/0000-0002-3852-3590}{\includegraphics[scale=0.5]{orcid.jpg}}}$\thanks{NASA Sagan Fellow}
\\
$^{1}$ School of Physics \& Astronomy, University of Birmingham, Edgbaston, Birmingham, B15 2TT, UK \\
$^{2}$ Center for Computational Astrophysics, Flatiron Institute, New York, NY 10010, USA \\
$^{3}$ Department of Physics and Astronomy, University of Exeter, Exeter, EX4 4QL, UK \\
$^{4}$ DTU Space, Technical University of Denmark, Elektrovej 328, DK-2800 Kgs. Lyngby, Denmark \\
$^{5}$ Instituto de Astrof{\'\i}sica de Canarias, E-38205 La Laguna, Tenerife, Spain \\
$^{6}$ Universidad de La Laguna, Dept. Astrof{\'\i}sica, E-38206 La Laguna, Tenerife, Spain \\
$^{7}$ Department of Physics, University of Oxford, Oxford OX1 3RH, UK \\
$^{8}$ Astrophysics Research Centre, School of Mathematics and Physics, Queen’s University Belfast, Belfast BT7 1NN, UK \\
$^{9}$ Subaru Telescope, National Astronomical Observatory of Japan (NAOJ), 650 N Aohoku Place, Hilo, HI 96720, USA \\
$^{10}$ Hamburger Sternwarte, Gojenbergsweg 112, D-21029 Hamburg, Germany \\
$^{11}$  Landessternwarte, Zentrum f\"ur Astronomie der Universit\"at Heidelberg, 69118 Heidelberg, Germany \\
$^{12}$ Astrophysics \& Space Institute, Schmidt Sciences, New York, NY 10011, USA \\
$^{13}$ Instituto de Astrof\'{\i}sica e Ci\^encias do Espa\c co, Universidade do Porto, CAUP, Rua das Estrelas, 4150-762, Porto, Portugal \\
$^{14}$ Departamento de F\'{\i}sica e Astronomia, Faculdade de Ci\^encias, Universidade do Porto, Rua Campo Alegre, 4169-007, Porto, Portugal \\
$^{15}$ INAF -- Osservatorio Astrofisico di Arcetri, Largo E. Fermi 5, 50125 Firenze, Italy \\
$^{16}$ Department of Astronomy \& Astrophysics, University of Chicago, Chicago, IL, USA \\
}

\date{Accepted 2025 May 16. Received YYY; in original form ZZZ}

\pubyear{2025}

\begin{document}
\label{firstpage}
\pagerange{\pageref{firstpage}--\pageref{lastpage}}
\maketitle

\begin{abstract}
As the fields of stellar and exoplanetary study grow and revolutionary new detection instruments are created, it is imperative that a homogeneous, precise source of stellar parameters is available. This first work of the \texttt{gr8stars} collaboration presents the all-sky magnitude limited sample of 5645 bright FGKM dwarfs, along with homogeneously derived spectroscopic parameters of a subset of 1716 targets visible from the Northern hemisphere.
We have collected high-resolution archival and new spectra from several instruments. 
Spectrosocpic parameters are determined using the PAWS pipeline, employing both the curve-of-growth equivalent width method, and the spectral synthesis method. We achieve median uncertainties of 106K in stellar effective temperature, 0.08 dex in surface gravity, and 0.03 dex in metallicity. This paper also presents photometric stellar parameters for these dwarfs, determined using SED fitting. The full \texttt{gr8stars} sample selection, including derived spectroscopic and photometric parameters, is made available through an interactive online database. We also perform a kinematic analysis to classify these stars according to their Galactic component.  

\end{abstract}

\begin{keywords}
stars: atmospheres -- techniques: spectroscopic -- stars: solar type
\end{keywords}



\section{Introduction}
\label{intro}

Our desire to understand our solar system drives us to observe stars and planets with configurations similar to our own.  Observations of solar-like stars, here defined as FGK main-sequence stars,  open the way to understanding the solar system's formation and evolution, with investigations into orbiting planets solidifying and expanding such knowledge. In particular, observations of an `Earth twin' -- that is, an Earth-like planet orbiting a sun-like star with an orbital period similar to Earth's -- would be a crucial cornerstone in this study. The search for these long period, small exoplanets will rely on the next generation of transit \citep[e.g., PLATO;][]{rauer2014} and extreme precision radial velocity (EPRV) surveys (e.g., the Terra Hunting Experiment \citep{hall2018}, the NEID Earth Twin Survey \citep{gupta2021, Gupta2025}).

A homogeneously characterised input sample for EPRV surveys is crucial for target selection \citep[e.g.,][]{crass2021, hojjatpanah2019}. Target selection for such a survey is rather complex, with many potential selection criteria depending on information that can be obtained from high-resolution stellar spectra. Impactful stellar properties gained from these spectra such as metallicity, magnitude, spectral type, and rotational broadening will affect the intrinsic radial velocity (RV) information from the star, so learning these properties will allow us to gauge detection limits for exoplanetary searches. High-resolution spectra additionally allow the investigation of binarity and stellar activity, which further complicate the extraction of planetary signals.

Stellar activity is a particularly difficult obstacle in its obscuration of planetary RV signals: an Earth twin will produce RV amplitudes on the order of $\rm 10\,cm\,s^{-1}$, which are readily overwhelmed by the amplitude of stellar activity signals \citep[e.g.,][]{Robertson2014, meunier2015, Fischer2016, suarezmascareno2017, zicher2022, almoulla2023}. It is therefore understandable why no such Earth twin has yet been observed. 

The challenges presented to the search for an Earth twin have led to the creation of the Terra-Hunting Experiment (THE) \citep{hall2018}, which will draw upon the upcoming HARPS-3 instrument on the Isaac Newton Telescope {(INT)\footnote{\url{https://www.ing.iac.es//astronomy/telescopes/int/}}} \citep{thompson2016}. The orbital periods of Earth twins require a long survey to repeatedly observe the RV signal produced. The THE will therefore require a select few stars for repeat observations across a period of many years. Such targets must be solar-like stars, bright enough to achieve high signal-to-noise ratio (SNR) observations, and observable from the INT site in La Palma. In this paper, we introduce the \texttt{gr8stars} catalogue, a key foundation in the target selection of the THE, and other long-term RV surveys. 

\texttt{gr8stars} was created as a collaborative effort to ensure the preparedness of the exoplanetary field for upcoming EPRV missions and studies. The main products of \texttt{gr8stars} are the library of homogeneously derived stellar parameters for the included stars, in addition to the database of uniformly-formatted spectra in both one-dimensional, and order-by-order formats. Aside from the primary aim of EPRV preparations, the collaboration and resulting catalogue will benefit many other science cases. For example, exoplanet occurrence rates and properties can be studied as a function of stellar properties occurrence using \texttt{gr8stars} as a homogeneous source of stellar parameters to investigate planet formation \citep[e.g.,][]{mortier2013, buchhave2014, adibekyan2021}. Target selection using \texttt{gr8stars} additionally does not have to be limited to EPRV surveys; direct imaging missions and planetary atmosphere studies, among many others, will be able to draw upon the information provided by the catalogue. The catalogue will also aid in the identification of solar twin stars, which are crucial to the understanding of stellar evolution and development of stellar atmospheric models. 

Looking to the future, it is anticipated that many of the bright, solar-like stars in \texttt{gr8stars} will benefit from asteroseismology observations with PLATO \citep{rauer2014}. The homogeneous stellar effective temperatures and metallicities from \texttt{gr8stars} will benefit the homogeneity of asteroseismic masses, radii, and ages for such stars. Additionally, the high-resolution, high SNR spectra used as part of \texttt{gr8stars} provide an excellent dataset for stellar abundance studies. Such investigations are crucial for not only the understanding of galactic chemical evolution \citep[e.g.,][]{mcwilliam1997}, but also benefit studies of lithium enrichment \citep[e.g.,][]{tsantaki2023} and planetary composition \citep[e.g.,][]{santos2017}.

In Section \ref{gr8} we describe the \texttt{gr8stars} catalogue and its context in the fields of stellar and exoplanetary physics. The target selection for the catalogue is described in terms of the selection criteria used, in addition to plans for the expansion of the catalogue. The structure of the catalogue and the spectra it contains are also detailed. Section \ref{data} describes the observations used to create the catalogue and the instruments used to conduct them. The methods used to determine stellar parameters for the targets within the catalogue are described in Section \ref{method}. Section \ref{results} then displays the results achieved, with a concluding discussion of these in Section \ref{conclusion}.

\section{The gr8stars catalogue}
\label{gr8}
The \texttt{gr8stars} catalogue provides an easy-to-access source of homogeneously prepared one-dimensional spectra in the lab frame, in addition to the full complement of \'echelle order-by-order spectra available for RV studies. Once completed using facilities from both hemispheres, the catalogue will include 5645 FGKM main-sequence stars. Homogenous stellar atmospheric parameters -- namely effective temperature ($T_{\rm eff}$), surface gravity ($\log\,g$), metallicity ($\rm [Fe/H]$), microturbulent velocity ($v_{\rm mic}$), macroturbulent velocity ($v_{\rm mac}$), and projected rotational velocity ($v \sin i_{\star}$) -- are provided for all targets with available spectra, as described throughout this paper. With the combination of stellar parameters and downloadable high-resolution spectra, the \texttt{gr8stars} catalogue provides a robust, reliable, and extensive source of materials and information for future studies. Amongst undoubtedly many others, the areas of exoplanet characterisation, precision stellar abundances, and ultimately direct imaging of exoplanetary systems can benefit from the resources provided by this catalogue.

The \texttt{gr8stars} catalogue focuses on bright FGKM stars. We have begun with stars observable from the Northern Hemisphere, with plans to expand this to the Southern Hemisphere in the future. Target selection was based on the Gaia DR3 catalogue \citep{gaia2016, gaia2021}, which provides crucial measurements for 1.8 billion sources including their colour, brightness, and parallax. 

\begin{figure}
    \centering
    \includegraphics[width=0.5\textwidth]{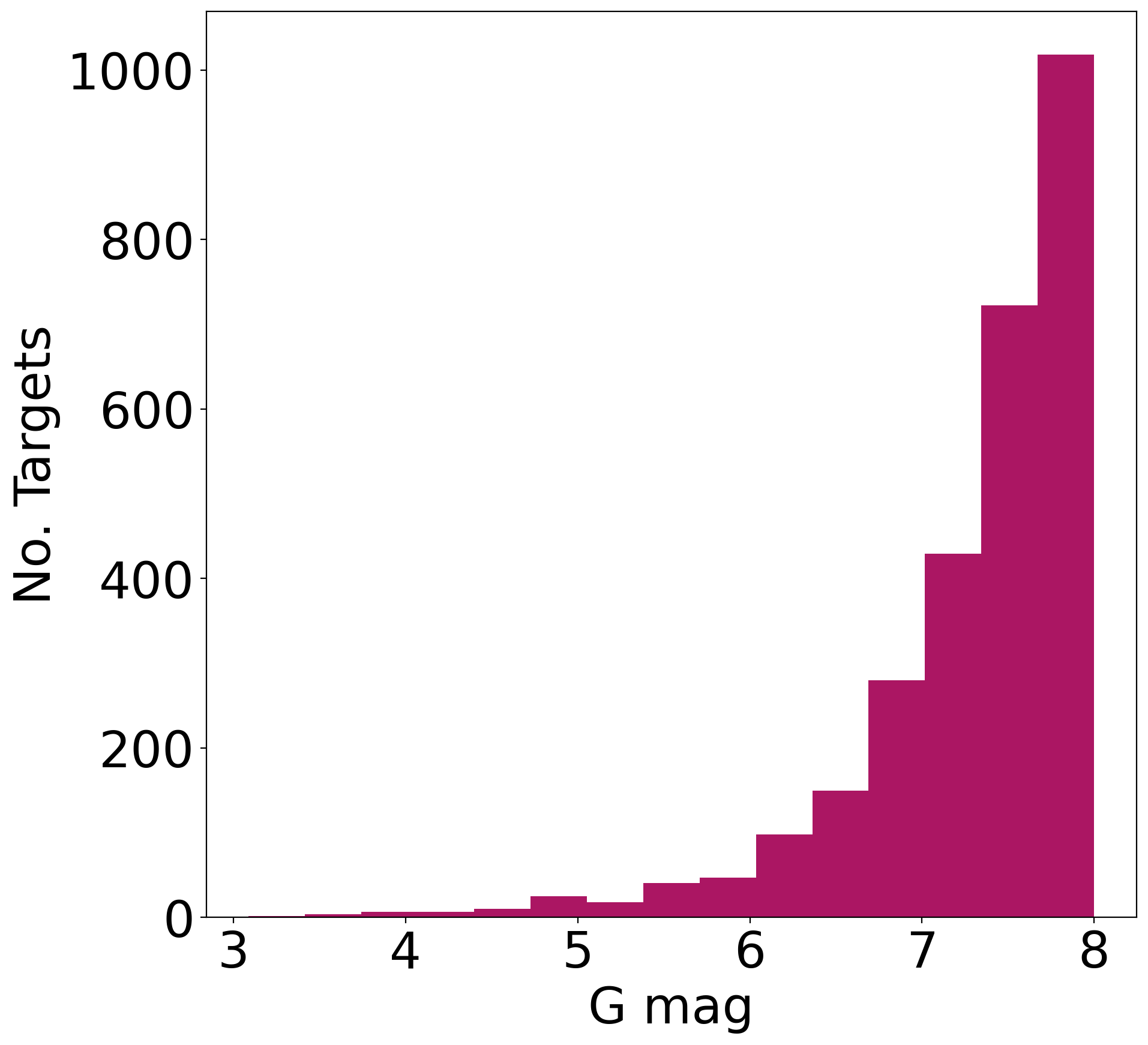}
    \caption{Histogram of apparent G magnitudes for the \texttt{gr8stars} targets}
    \label{fig:mag_hist}
\end{figure}

Broadly, the selected targets for the catalogue were determined by four main requirements. Firstly, stars must be brighter than an apparent Gaia G-band magnitude of 8, ensuring that high signal-to-noise (SNR) observations (ideally SNR $\ge$ 200, \citep{sousa2008}) could be available for all targets. Photometric and astrometric Gaia measurements were used to calculate the absolute G magnitudes and extract the BP$-$RP colours. Given the focus of the catalogue on bright, nearby stars, extinction was assumed to be small here, hence parallax-corrected G magnitude could be used as a proxy for absolute G magnitude. The distribution of the magnitudes of the \texttt{gr8stars} targets is shown in Figure \ref{fig:mag_hist}.  It is worth noting that there is a slight lack of completeness for extremely bright targets in Gaia eDR3 due to saturation ($\approx$ 20\% of $\rm{G} \le$ 3 targets do not have entries \citep{gaia2016, gaia2021b, gaia2021c}). However, in addition to the majority of these being unsuitable for \texttt{gr8stars} due to their nature as giant stars, we anticipate that future work within the \texttt{gr8stars} collaboration will require solutions from Gaia DR3, hence ensuring we only use targets with complete Gaia information helps to preserve completeness for future papers.

Focusing on FGKM main-sequence stars required us to cut all earlier type main-sequence stars, in addition to giants and subgiants. The majority of such sources were removed by the implementation of the minimum Gaia BP$-$RP colour $\ge 0.6$, where the lack of lower limit allows for the inclusion of bright M dwarfs.

\begin{figure*}
    \includegraphics[width = \textwidth]{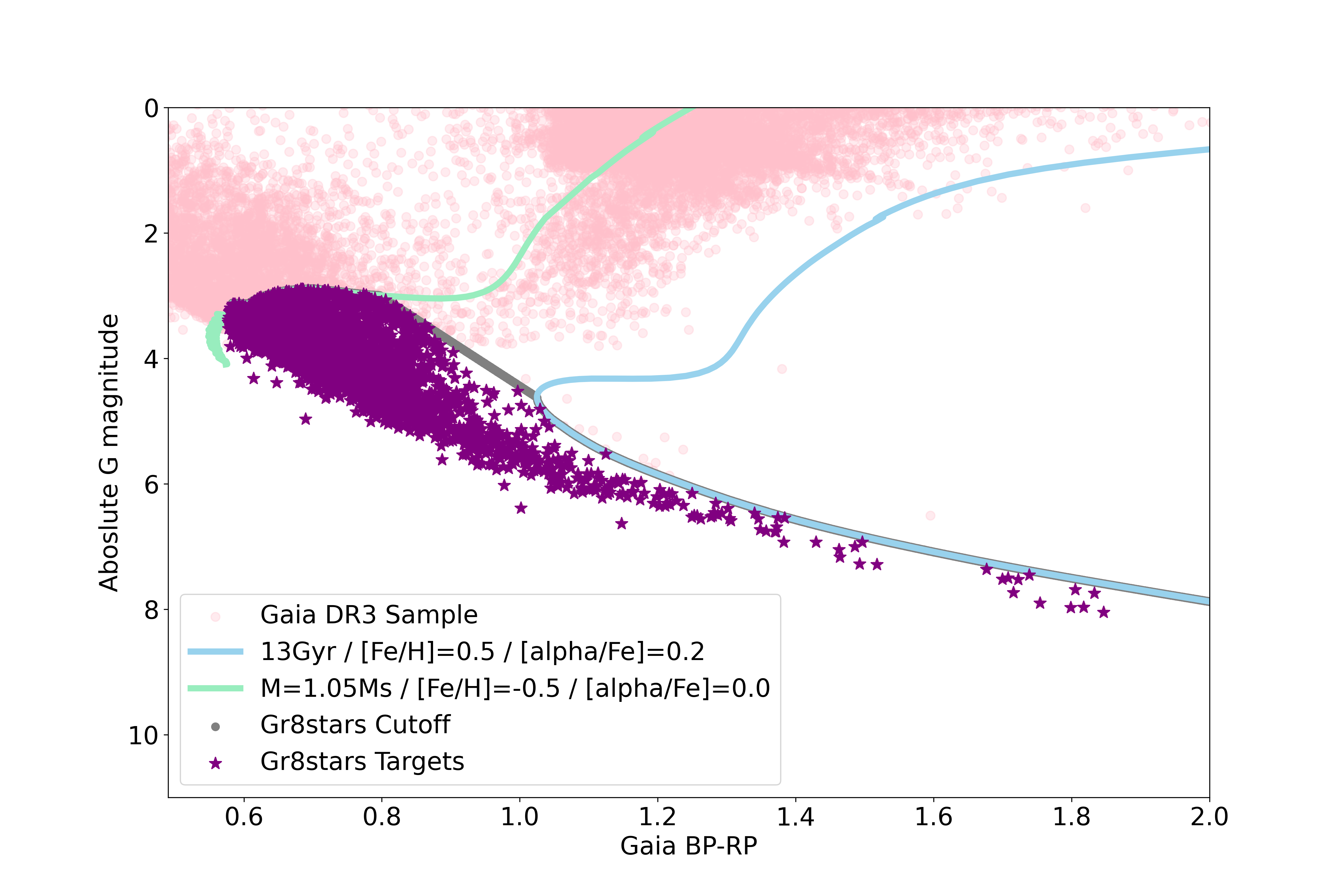}
    \caption{The isochronal and mass track cuts used in the target selection. The 13 Gyr, metal rich ($\rm [Fe/H]$ = 0.5) isochrone, and the metal poor ($\rm [Fe/H]$ = -0.5), 1.05 Solar Mass (M$_{\odot}$) mass-track used in selection are shown in green and blue, respectively. Targets meeting these two cuts are shown in purple, with targets not selected shown in light pink.}
    \label{isochrones}
\end{figure*}

Evolved stars were removed from the sample by the use of isochrones. Figure \ref{isochrones} displays the isochronal cuts made on the sample, using the Dartmouth isochrones and evolutionary tracks \citep{Dotter2008}. An iron-poor ([Fe/H] = $-$0.5) mass track of 1.05 M$_{\odot}$ ensured the inclusion of metal-poor G dwarfs, and a 13 Gyr isochrone for iron-rich ([Fe/H] = 0.5) sources was used to define the upper edge of the main sequence.

Observability from northern hemisphere instruments was desired for all targets. To ensure high-quality observations with minimal airmass, we selected only targets with a declination larger than or equal to $-15\degree$. 

\begin{figure*}
    \includegraphics[width=\textwidth]{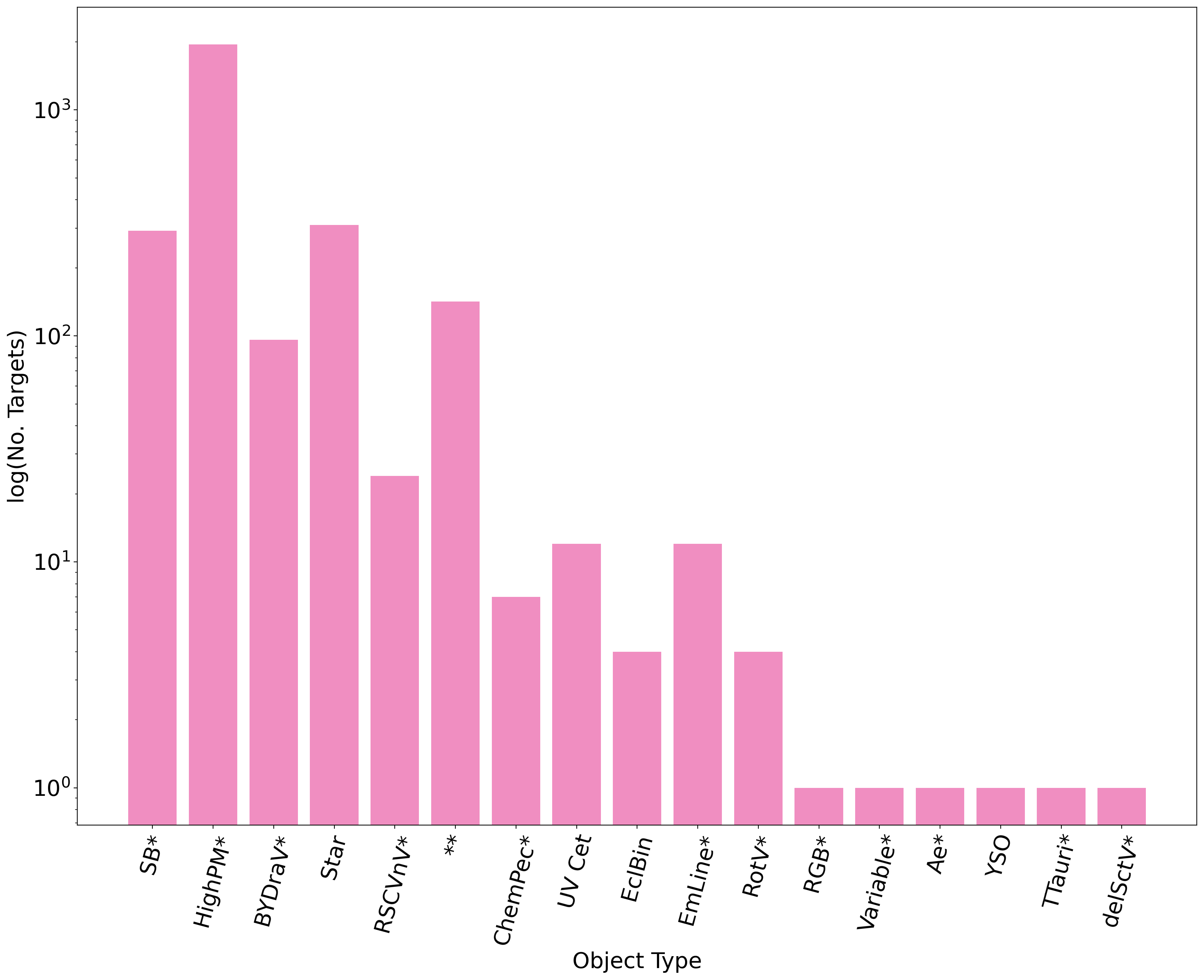}
    \caption{
    A histogram representing the number of targets belonging to each SIMBAD object type within the \texttt{gr8stars} catalogue.  Table \ref{tab:simbads} contains definitions for the SIMBAD object types.
    }
    \label{star_types}
\end{figure*}

The diversity of the targets within the \texttt{gr8stars} sample is displayed as a series of histograms in Figure \ref{star_types} according to their SIMBAD object types \citep{wenger2000}. Unsurprisingly, the vast majority of targets belong to either the HighPM* (High Proper Motion Star) or Star categories, with these being `regular' single stars, however many also fit into other regimes. Many types of variable stars are seen within the sample, with the majority being BY Dra and RS CVn variables, in addition to both spectroscopic and eclipsing binaries. In the case where the main SIMBAD object type is listed as an Eruptive Variable, we use the classification in this work as UV Cet. As the term Eruptive Variable encompasses a broad definition of stars, with many mechanisms of stellar variability, we have traced all of our targets under this classification to be specifically UV Ceti Variables, classified by \citet{gershberg1999}. As the \texttt{gr8stars} catalogue is intended to be a foundation for a variety of studies, targets of all object types are kept in the sample, to attempt to minimise the degree of bias present in the sample.

After selection criteria were enforced, the final sample size for Northern-hemisphere observable targets is 2858 targets. Of these, we have obtained either archival or new observations for 1716 targets. Considering a brighter subsample of G < 7.5, which may be more desirable for EPRV surveys, the completeness of \texttt{gr8stars} is 986 of 1418 targets.

\section{Data}
\label{data}
Spectra used to form the \texttt{gr8stars} catalogue were taken using the FEROS, FIES, HARPS, HERMES, SOPHIE, and UVES high-resolution, optical spectrographs. The statistics of the observations made are shown in Table \ref{tabobs_stats}, where resolution and wavelength coverage for the observation modes used in \texttt{gr8stars} are also specified. A total of 36019 spectra were used to create the final \texttt{gr8stars} products, with the bulk of observations conducted using the SOPHIE spectrograph.

\begin{table*}
    \centering
    \caption{Observation statistics from the instruments used in the \texttt{gr8stars} catalogue, with spectral resolution and wavelength coverage specified for the instrumental setups used as part of \texttt{gr8stars}, rather than representing all possible configurations. The number of overlapping targets that have been observed by multiple spectrographs is also displayed here.}
    \begin{tabular}{cccccc}
        Spectrograph & No. Targets  & Overlap & Mean SNR & Resolving Power & Wavelength Coverage\\
        \hline
         FEROS & 18 & UVES 6, HARPS 1, SOPHIE 1 & 299 & 48 000 & 350 - 920 nm\\
         FIES & 139 & HERMES 8, SOPHIE 1& 89 & 67 000 & 365 - 730 nm \\
         HARPS & 375 & SOPHIE 161, FEROS 1, HERMES 1, UVES 8 & 278 & 115 000 & 378 - 691 nm \\
         HERMES & 54  & FIES 8, HARPS 1, SOPHIE 3& 165 & 85 000 & 377 - 900 nm  \\
         SOPHIE HR &  1286 & FIES 1, HARPS 161, FEROS 1, HERMES 3, UVES 2 & 196 & 75 000 & 387.2 - 694.3 nm\\
         UVES & 42 & FEROS 6, HARPS 8, SOPHIE 2 & 244 &  80 000 & 320 - 1050 nm\\
    \end{tabular}
    
    \label{tabobs_stats}
\end{table*}

A blend of both archival observations and new observing programmes were used to create the \texttt{gr8stars} catalogue. The varying properties of each instrument, described in the following subsections, results in systematic differences in the spectra available. To minimise the effects of these, the spectra are all processed homogeneously after the respective instrumental reduction pipelines to produce the combined spectra available as part of the \texttt{gr8stars} catalogue.

\subsection{FEROS}
22 archival spectra were used from the FEROS (Fiber-fed Extended Range Optical Spectrograph) spectrograph \citep{Kaufer1999} as part of the \texttt{gr8stars} catalogue. These spectra were taken between 2004 and 2022, with the program IDs listed in Appendix \ref{progids}. Located at the  La Silla Obvservatory, FEROS is mounted on the MPG/ESO 2.2\,m telescope and achieves a resolving power of $R \approx 48 000$, with a wavelength range of 350 - 920 nm. The archival spectra were accessed as pre-processed science spectra, having been reduced by the FEROS data reduction software (DRS), implemented in ESO-MIDAS \citep{ESO2013}. The FEROS DRS is applied in real-time as observations are taken.

\subsection{FIES}
The FIES (FIbre-fed Echelle Spectrograph) spectrograph \citep{Telting2014} is located on the Nordic Optical Telescope (NOT) in La Palma, with a wavelength coverage of 365 - 730 nm, and a resolving power of $R \approx 67 000$ using the high-res fibre. A combination of archival observations and new observations from 2017 to 2023 were used for 139 targets, supplied as pre-merged S1D spectra using FIEStool\footnote{\url{https://www.not.iac.es/instruments/fies/fiestool/FIEStool.html}}. New observations were conducted under Proposal ID 66-010 (Gr8stars: A survey of the brightest Northern FGKM dwarf stars; PI: Bucchhave); Proposal IDs for the archival spectra are shown in Appendix \ref{progids}.

\subsection{HARPS}
Mounted on the ESO 3.6m telescope at La Silla Observatory, the HARPS (High Accuracy Radial velocity Planet Searcher) spectrograph \citep{Mayor2003} has a wavelength range of 378 - 691 nm and resolving power of $R \approx 115 000$. 375 targets were observed using HARPS between 2003 and 2019, totalling 12526 individual spectra, all reduced by the HARPS DRS \citep{lovis2007}. The programme IDs for the observations can be found in Appendix \ref{progids}.

\subsection{HERMES}
The HERMES (High-Efficiency and high-Resolution Mercator Echelle Spectrograph) spectrograph \citep{Raskin2011} operates on the 1.2m Mercator telescope in La Palma, having a spectral range of 377 to 900 nm, and a resolving power of $R \approx 85 000$. 54 targets were observed using HERMES, with spectra supplied as a single combined spectrum for each target by the HERMES DRS \citep{Raskin2011}. Observations were taken as part of the new programme 123-Mercator4-A/20B (PI: Mortier) in 2021. 

\subsection{SOPHIE}
Owing to their northern observability, the majority of the \texttt{gr8stars} targets were observed using the SOPHIE (Spectrographe pour l’Observation des Ph\'enom\'enes des Intérieurs stellaires et des Exoplan\`etes) spectrograph \citep{bouchy2009, SOPHIE}, mounted on the 1.93 m reflector telescope at the Haute-Provence Observatory. Two observation modes are available using SOPHIE -- for the \texttt{gr8stars} targets, we used only HR (High Resolution) mode observations, achieving a resolving power of $R \approx 75 000$ across the wavelength range of 387.2 nm to 694.3 nm. 1286 targets were observed using SOPHIE between 2006 and 2020, totalling 22 720 individual spectra with programme IDs specified in Appendix \ref{progids}. The spectra were all processed by the SOPHIE DRS \citep{bouchy2009}.

\subsection{UVES}
Archival spectra for 42 \texttt{gr8stars} targets were obtained from the UVES (Ultraviolet and Visual Echelle Spectrograph) spectrograph \citep{Dekker2000} on the VLT Kueyen telescope at Paranal. UVES has a wavelength range of 320 to 1050 nm, able to achieve resolving powers of $R \approx 80 000$ and $R \approx 110 000$ dependent on observation mode. UVES observations were used for 27 targets, with a total of 130 spectra reduced by the UVES DRS \citep{ballester2000} between 2020 and 2022. A mix of red and blue arm observations were available, hence varying resolutions of UVES spectra are used in this analysis, and made available as part of the \texttt{gr8stars} catalogue.

The programme IDs for the observations can be found in Appendix \ref{progids}.

\subsection{Spectral Format}
A uniform, non-continuum-normalised format is applied to all spectra in the catalogue, allowing for ease of use and automation of analysis.  Spectra are named according to the form `X\_INST\_SYD.fits', where X is the target's primary name (selected as the SIMBAD main identifier), INST refers to the instrument from which the spectrum originates, and Y is either 1 or 2, depending on whether the spectrum is one- or two-dimensional, respectively. The S1D spectra, which are formatted from the S1D product produced by each instrument's reduction pipeline, contain two FITS extensions. The first extension (i.e., index 0) contains the flux values as a data product, and the header contains the start wavelength and fixed wavelength step, allowing the spectrum to be built solely from this extension. In the case that the original S1D spectrum is not sampled uniformly, it was resampled uniformly to form the \texttt{gr8stars} S1D product.  The header for this extension also contains the target information displayed in Table \ref{tab:header}. The second extension (index 1) contains a table of the wavelength, flux, and flux errors, again allowing the spectrum to be built from this. Having both formats of information available allows for a versatile array of methods to be used for spectral analysis. Where available, S1D spectra from the same instrument are combined for each target to produce a higher SNR final product, with the individual spectra files used specified in the index 0 header. In cases where wavelength gaps are present in the spectra, flux values are set to \texttt{NaN}. By nature of being combined from multiple exposures, the S1D spectra have all been RV corrected to the lab frame, where the RV was determined using a cross-correlation function (CCF) and G2 mask.

The order-by-order S2D spectra for all exposures are also individually available to allow for customised reduction pipelines and analysis methods, following the same naming convention. The files comprise of 3 FITS extensions, the first (index 0) containing the fluxes, the second (index 1) containing the blaze corrections, and the third (index 2) being the wavelengths, with all three extensions containing arrays for the separate orders. 

\section{Stellar Parameter Determination}
\label{method}
As part of the \texttt{gr8stars} catalogue, we make available stellar parameters homogenously derived from both spectroscopic and photometric methods. The parameters derived by these methods complement each other well, with common parameters allowing for direct comparison, and those exclusive to each method adding to the completeness of our knowledge of each star.

\subsection{Spectroscopic Methods}
\label{spec_method}
To determine the atmospheric parameters for the \texttt{gr8stars} targets using their spectra, we made use of the \texttt{PAWS} pipeline as described by \citet{Freckelton2024}, which employs the functionalities of \texttt{iSpec} \citep{blancocuaresma2014, blancocuaresma2019}. The pipeline is presented as an easy-to-use GUI, combining two methods, both using the ATLAS9 set of model atmospheres \citep{kurucz2005}, to produce reliable atmospheric stellar parameters. The \texttt{gr8stars} spectra are supplied as coadded spectra in the lab frame, hence many of the spectral preparation steps in the pipeline were omitted. The spectra were normalised by modelling the continuum using a B-spline of 2 degrees fitted every 5\,nm, and then dividing the flux at each point in the spectrum by the fitted continuum flux.

Once spectra were prepared as described above, the equivalent width (EW) method was employed to generate a set of initial stellar parameters ($T_{\rm eff}$, $\log\,g$, $\rm [Fe/H]$, and $v_{\rm mic}$) using the \texttt{WIDTH} \citep{width} radiative transfer code to ensure ionisation and excitation balance in the Fe I and Fe II lines. These parameters were then fed as inputs to the spectral synthesis method (using the \texttt{SPECTRUM} \citep{spectrum} radiative transfer code), which derived the final parameters - $T_{\rm eff}$, $\log\,g$, $\rm [Fe/H]$,$v_{\rm mic}$, $v_{\rm mac}$, and $v \sin i_{\star}$. The spectral synthesis method is applied in the wavelength region 480 - 680 nm, using the \texttt{SPECTRUM} line list based on the NIST atomic database \citep{ralchenko2005}. Each of the parameters, aside from macroturbulent velocity ($v_{\rm mac}$ which was derived from the $T_{\rm eff}$, $\log\,g$, and $\rm [Fe/H]$ using an empirical relation from the results of the Gaia-ESO survey \citep{jofre2014, blancocuaresma2014}), is a free parameter in at least one stage of the pipeline, allowing for as little model dependence as possible.  As described by \citet{blancocuaresma2014}, errors on the determined parameters are derived from the covariance matrix resulting from least-squares fitting. As these uncertainties reflect precision only, we account for model accuracies by adding uncertainties on $T_{\rm eff}$ by 100 K in quadrature, as is standard for PAWS \citep{Freckelton2024}. Additionally, we follow \citet{sousa2011}, adding accuracy uncertainties on $\rm [Fe/H]$ and $\log\,g$ of 0.04 and 0.1 dex in quadrature respectively.  PAWS can be applied to all FGKM main-sequence stars with high-resolution, high SNR spectra, hence was used for the whole \texttt{gr8stars} dataset.

\subsection{Spectral Energy Distribution Fitting}
\label{SED_fitting}
In addition to the spectroscopic analysis, we computed the stellar radius $R_\star$ and effective temperature $T_{\rm SED}$ (to differentiate it from spectroscopically derived measures of temperature) using a spectral energy distribution (SED) fitting method. 
We first applied this method to low-mass stars in \citet{morrellExploringMdwarfLuminosityTemperatureRadius2019} and \cite{Morrell:2020aa}, with \citet{morrellUsingStellarSpectral} demonstrating that the technique generalises to Solar-like exoplanet hosts.

In contrast to spectroscopic analyses, which measure the characteristic of temperature dependent spectral features, our SED fitting method compares broadband, multi-colour photometry with synthetic photometry. 
Synthetic photometry were generated from the BT-SETTL Cosmological Impact of the First STars (CIFIST) \citep{Allard2012a} stellar atmosphere grid, diluted using the geometric distances of \citet{BailerJones2021a}.
By best matching the integrated area beneath and the overall shape of the SED, we determined the luminosity $L_{\rm SED}$ and temperature $T_{\rm SED}$, respectively.
Both parameters together unambiguously define $R_\star$.
Note that unlike the other analyses in this paper, this method self-consistently measures both $T_{\rm SED}$ and $R_\star$ using only photometry and distances, providing an alternate, orthogonal measure of temperature to spectroscopic methods. 

The photometric fitting was performed using the $G_{\rm BP}$ and $G_{\rm RP}$ bands from Gaia DR3  \citep{gaia2016, gaia2023}, the \textit{J}, \textit{H}, and \textit{K} bands from 2MASS \citep{Skrutskie2006}, and the \textit{W1}, \textit{W2}, and \textit{W3} bands from AllWISE \citep{Wright2010}.
For inclusion in the final catalogue we required that targets were fitted with 5 or more photometric bands, and the fit had $\chi^2<10$.
We calculated our statistical uncertainties following the methods described in \citet{morrellExploringMdwarfLuminosityTemperatureRadius2019}, \cite{Morrell:2020aa} and \citet{morrellUsingStellarSpectral}, adopting a floor value for the uncertainty in each photometric band of 0.01 mag, corresponding to an uncertainty of about 1 per cent.
We imposed this floor to both avoid the systematics present in Gaia photometry at below this level, and to down-weight the Gaia bands during the fitting process.

Although nearby, the target stars are still subjected to some extinction, which makes the observed SED appear cooler and less luminous than the star actually is. 
We compensated for this by fitting each target with a grid of atmosphere models reddened with the \citet{fitzpatrickAnalysisShapesInterstellar2019} extinction law. 
The amount of reddening applied to each target was determined by using the {\it Gaia} positions and parallaxes to place it within the 10\ pc resolution dust maps of \citet{vergelyThreedimensionalExtinctionMaps2022}.
We then integrated along the line of sight from the Sun to the target star to determine an extinction at 550\ nm ($A_{55}$). 
Combining this with the mean Milky Way total-to-selective extinction ($R(55)=3.02$), allows us to determine the nominal colour excess ($E(44 - 55)$) to which we should redden the models. As the extinctions to these targets are so low, errors in its determination will have little impact on our results.
All the stars in our sample have $E(44 - 55)< 0.03$, compared with the error of $E(44 - 55)= 0.01$ required to introduce an error of 1\% in radius or $T_{\rm SED}$ \citep{morrellUsingStellarSpectral}.
Hence it would require an error of 30\% in the determination of the extinction to significantly affect our results, whereas the uncertainties in $E(44 - 55)$ given by \citet{vergelyThreedimensionalExtinctionMaps2022} are always less than 0.001 mags within 100\,pc of the Sun for their 50\,pc resolution maps.

\subsection{Galactic Velocities}
\label{galacy_vels}
We determined the galactic velocities -- U,V,W -- in order to classify the galactic components of the \texttt{gr8stars} targets, with U being the velocity towards the galactic centre, V the velocity in the direction of the galactic rotation, and W the velocity towards the galactic north pole. These velocities were obtained following the matrix equations outlined by \citet{JohnsonSoderblom1987}, using the right ascension, declination, proper motions, parallax, and radial velocity from Gaia DR3 \citep{gaia2023}. Of the sample, 250 did not have an available radial velocity from Gaia DR3, hence were not included in this analysis.

\section{Results}
\label{results}
\subsection{Validation of Parameters}

\subsubsection{Atmospheric parameters}
\begin{figure}
    \centering
    \includegraphics[width=0.5\textwidth]{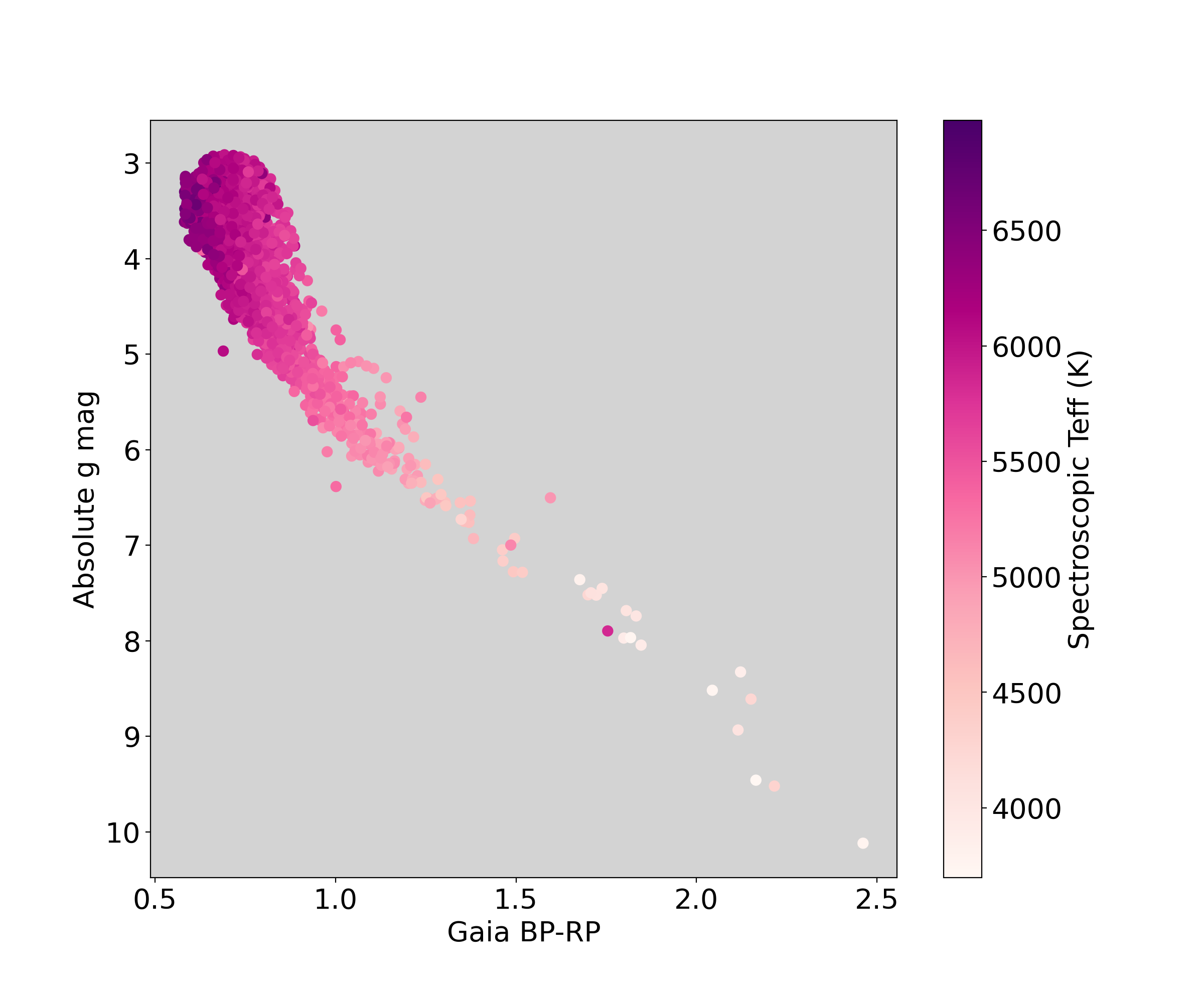}
    \caption{Colour-magnitude diagram of the \texttt{gr8stars} targets, with stars found to be hottest in this work as the darkest points, and those found to be the coolest as the lightest points.}
    \label{fig:hrdiagram}
\end{figure}

Given the breadth of the catalogue, results were expected to follow the trend of main-sequence stars in the solar neighbourhood with the exception of a small number of outliers. Figure \ref{fig:hrdiagram} displays a Gaia colour-magnitude diagram of the targets, with a colour scale representing the spectroscopic $T_{\rm eff}$ derived in this work. Overall, the targets follow the expectation of bluer, brighter stars being the hottest. A few outliers to this are present within the diagram, which we further investigate within this section.

\begin{figure}
    \centering
    \includegraphics[width=0.5\textwidth]{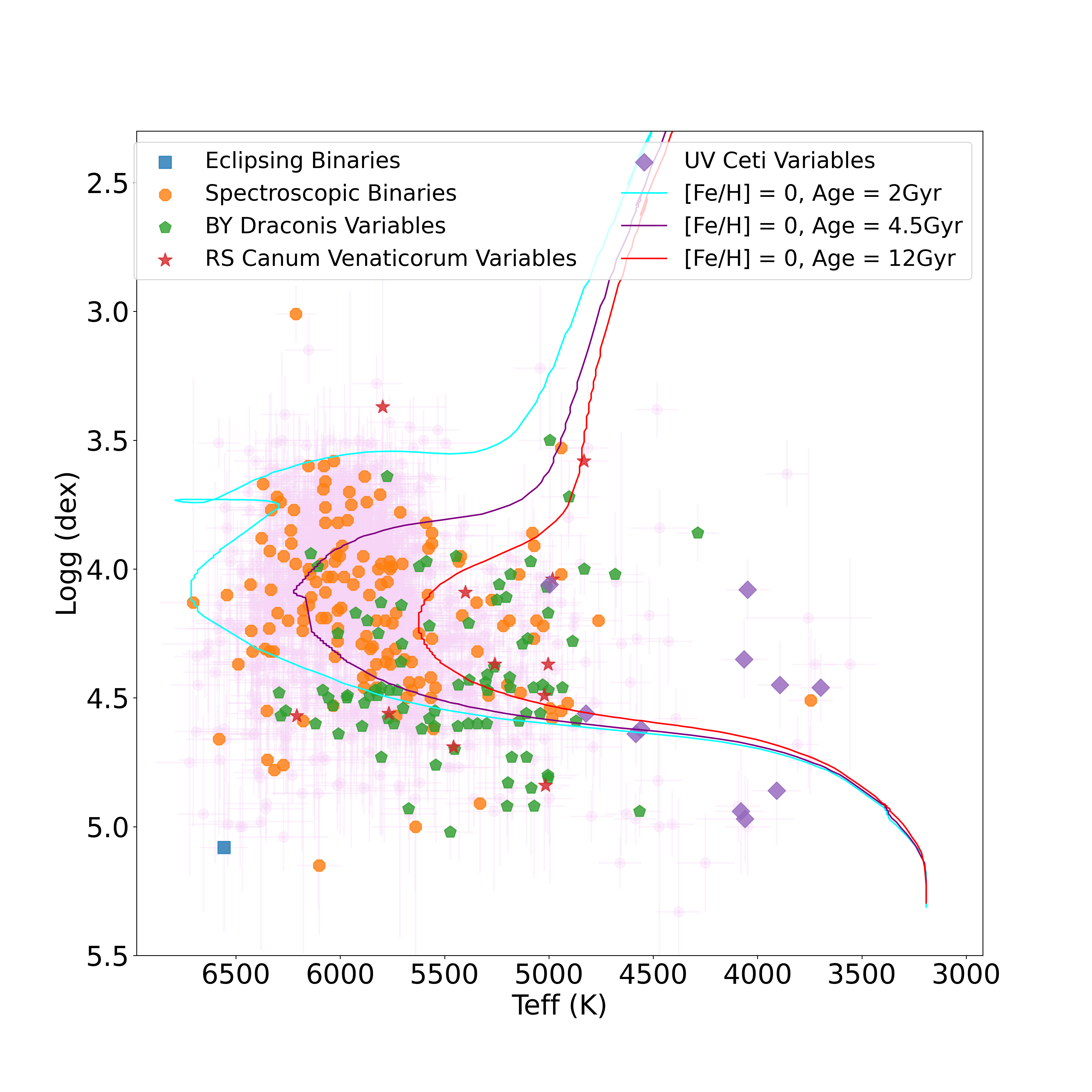}
    \caption{Spectroscopic surface gravity ($\log\,g$) and spectroscopic effective temperature ($T_{\rm eff}$) results from this work plotted against each other, with the legend identifying outliers of different categories. Dartmouth isochrones are over-plotted as solid lines \citep{Dotter2008}.}
    \label{fig:logg_teff_col}
\end{figure}

By cross-matching the \texttt{gr8stars} targets with the SIMBAD astronomical database\footnote{\url{https://simbad.u-strasbg.fr/simbad/sim-fid}} \citep{wenger2000}, all targets were classified into the 5 categories shown in Figure \ref{fig:logg_teff_col}. Each of these categories bring about challenges in the determination of atmospheric parameters from their stellar spectra, which may provide explanation for outliers in our results. Eclipsing binaries (EBs), spectroscopic binaries (SBs), and RS Canum Venaticorum variables (RS CVns) all consist of two stars, meaning the spectra from both stars can be `entangled' \citep[e.g.,][]{rawls2016,czekala2017,sairam2024}, in which the two spectra are blended. 

UV Ceti variables are typically M dwarfs that undergo stellar variability processes mainly characterised by short flares, causing brightness increases between 1 to 6 magnitudes \citep{gershberg1999}. RS CVn and BY Draconis (BY Dra) variables have prevalent chromospheric activity, resulting in large numbers of stellar spots. These spots may produce adverse effects in the stellar spectra from these stars, potentially resulting in inaccuracies in the atmospheric parameters derived. Additionally, in the context of the use of the \texttt{gr8stars} catalogue in selecting optimal targets for planet-searching surveys, high-activity stars may increase difficulty as outlined in Section \ref{intro}, hence it was important that we identify and highlight these within this work.

Figure \ref{fig:logg_teff_col} displays the \texttt{gr8stars} target spectroscopic results from this work in a $\log\,g$ vs $T_{\rm eff}$ diagram, with the outliers coloured according to the plot legend. With almost all targets being FGK main-sequence stars, the majority of points occupy the region 3.5 dex <  $\log\,g$ < 5.0 dex, 4500 K < $T_{\rm eff}$ < 6500 K. Several targets are cooler or hotter than this temperature range, however still show a derived $\log\,g$ appropriate for a main-sequence star, hence represent the extreme ranges of F and K type targets, extending into 3 M dwarfs present in the sample. More problematic, however, are those targets found to have $\log\,g$ < 3.5 dex. Such values are much lower than expectations for these stars, indicating that the spectra observed from these stars posed inherent difficulties for determining $\log\,g$, suggesting the value determined from SED fitting may be more suitable.  

\begin{figure}
    \centering
    \includegraphics[width = 0.5\textwidth]{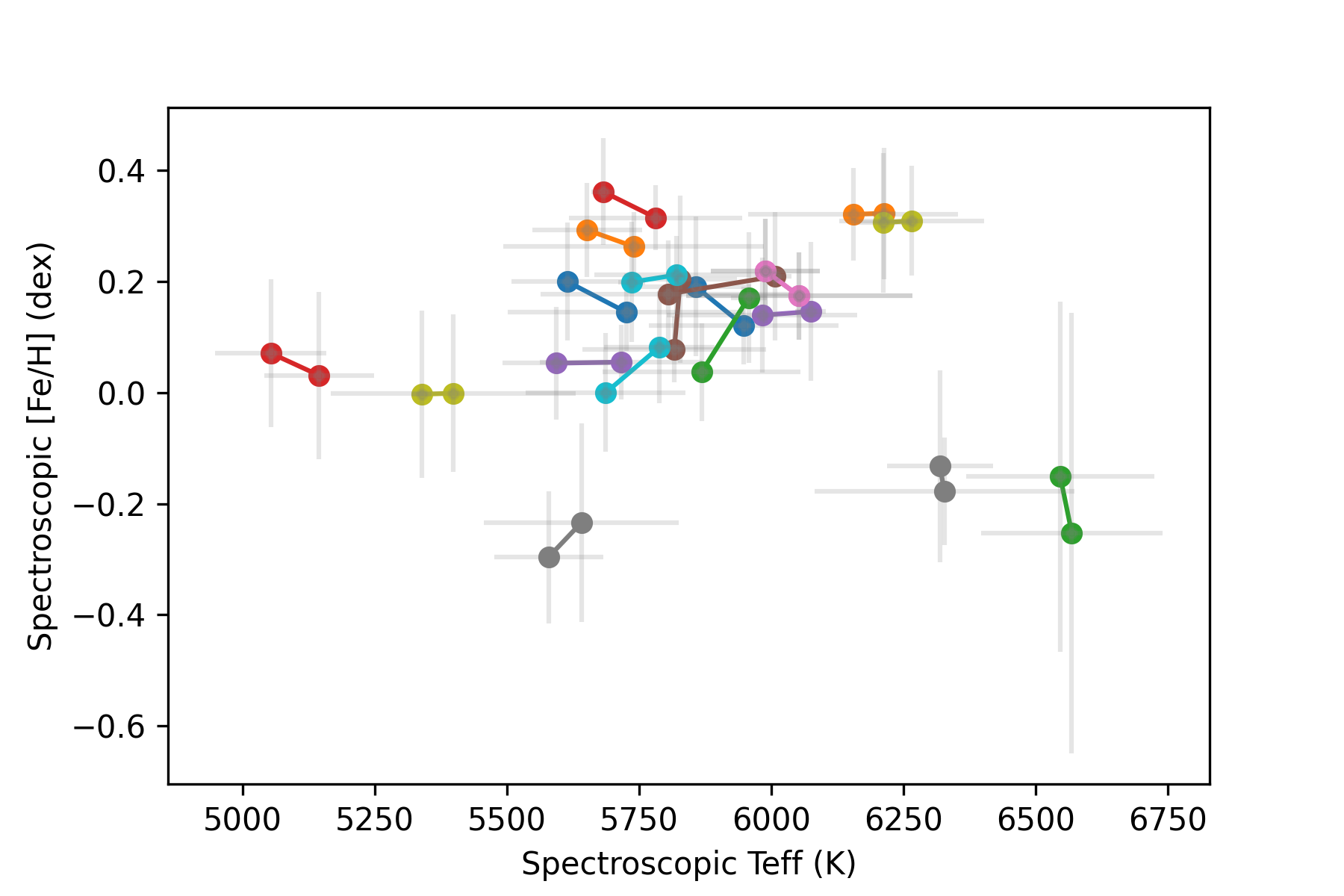}
    \caption{$ \rm [Fe/H]$ vs $T_{\rm eff}$, both derived in this work, for 20 randomly selected targets in the \texttt{gr8stars} catalogue. All 20 targets have spectra available form two different instruments, wherein which results for the same target from separate instruments are plotted in the same colour and joined by a line. Error bars for both measurements are shown in light grey.}
    \label{fig:inst_comp}
\end{figure}

To investigate the effects of any systematics introduced by the use of different instruments, 20 targets with observations from multiple spectrographs were selected for comparison.
Figure \ref{fig:inst_comp} displays the effective temperatures ($T_{\rm eff}$) and metallicities  ($\rm [Fe/H]$) determined for the 20 randomly selected targets.  From this, it is clear that the results from different instruments overall agree within uncertainties. Additionally, we note that Figure \ref{fig:inst_comp} does not appear to show any correlations between $ \rm [Fe/H]$ and $T_{\rm eff}$. Of the 20 targets plotted, 7 show $ \rm [Fe/H]$ to be increased for an increase in $T_{\rm eff}$, 8 show a decrease in $ \rm [Fe/H]$ for an increase in $T_{\rm eff}$, and 5 show no change in $ \rm [Fe/H]$. The mean difference between $T_{\rm eff}$ measurements for the same target is 78 K, with a standard deviation of 45 K; for the measured $\rm [Fe/H]$, the mean difference is 0.05 dex, having a standard deviation of 0.04 dex.

Using spectra from varying instruments is therefore not a concern with regard to the homogeneity of the \texttt{gr8stars} catalogue. However, we keep in mind that observations are all subject to environmental effects, and therefore low SNR observations may cause discrepancies in results from different instruments. In cases where multiple sources of spectra are available for a target, an inverse variance-weighted mean is taken for all parameters, with uncertainties on the final parameters taken from the variance of the input parameters. However, if observations from one instrument are deemed low quality (i.e., SNR < 50), results from only the higher-quality spectra are taken.

\begin{figure}
    \centering
    \includegraphics[width=0.5\textwidth]{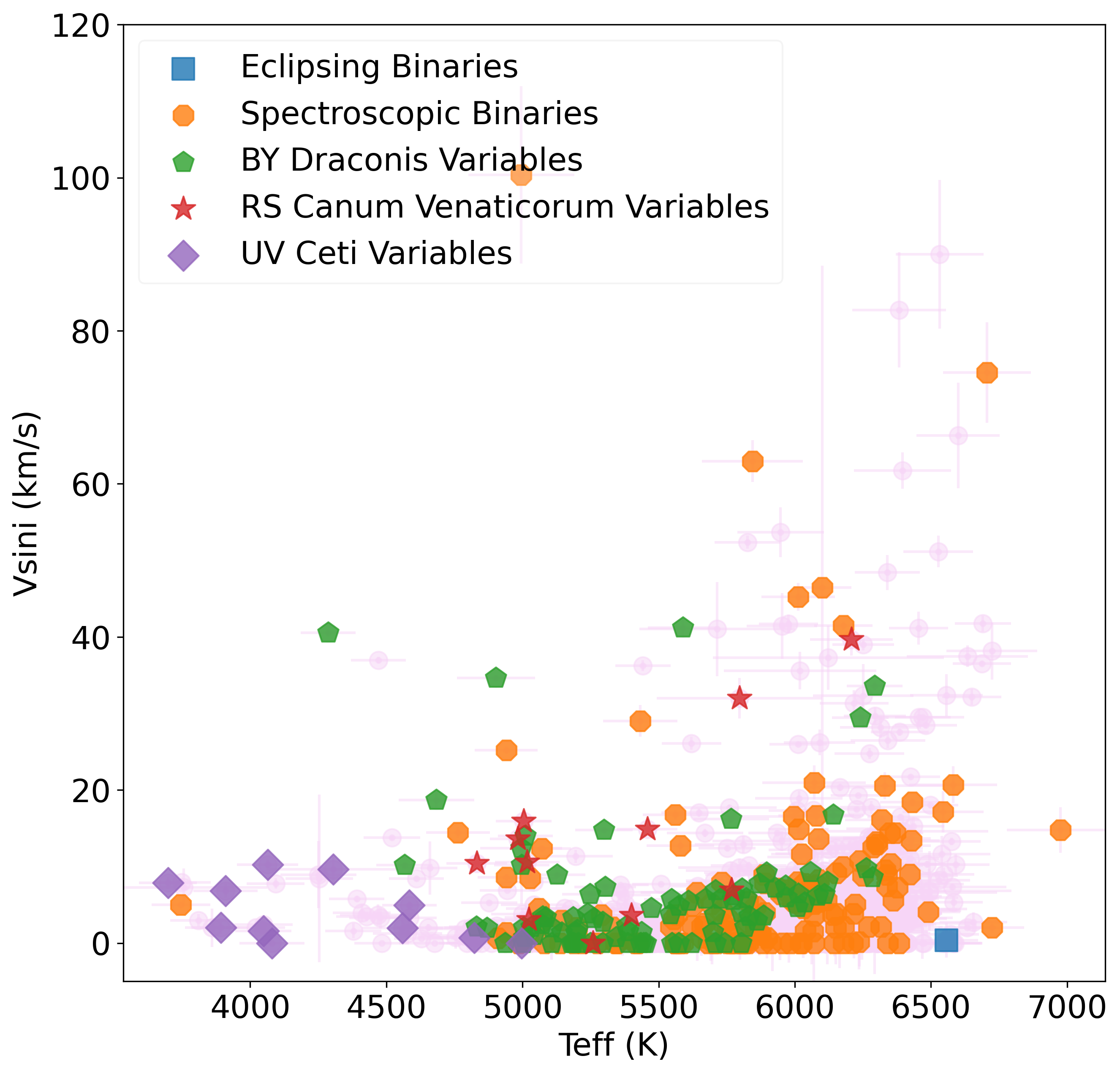}
    \caption{Spectroscopic projected rotational velocity ($v\sin i_{\star}$) plotted against the spectroscopic effective temperature ($T_{\rm eff}$) for the \texttt{gr8stars} targets, with outliers marked again as in Figure \ref{fig:logg_teff_col}.}
    \label{fig:vsini_teff}
\end{figure}

Outliers are further explored in Figure \ref{fig:vsini_teff}, in which the $v \sin i_{\star}$ is plotted against the spectroscopic $T_{\rm eff}$, both results from this work. Many of the higher $v \sin i_{\star}$ targets can be classified according to the outlier categories, indicating that the $v \sin i_{\star}$ measurement may have been inflated by broadening due to blended spectra.

\begin{figure}
    \centering
    \includegraphics[width=\columnwidth]{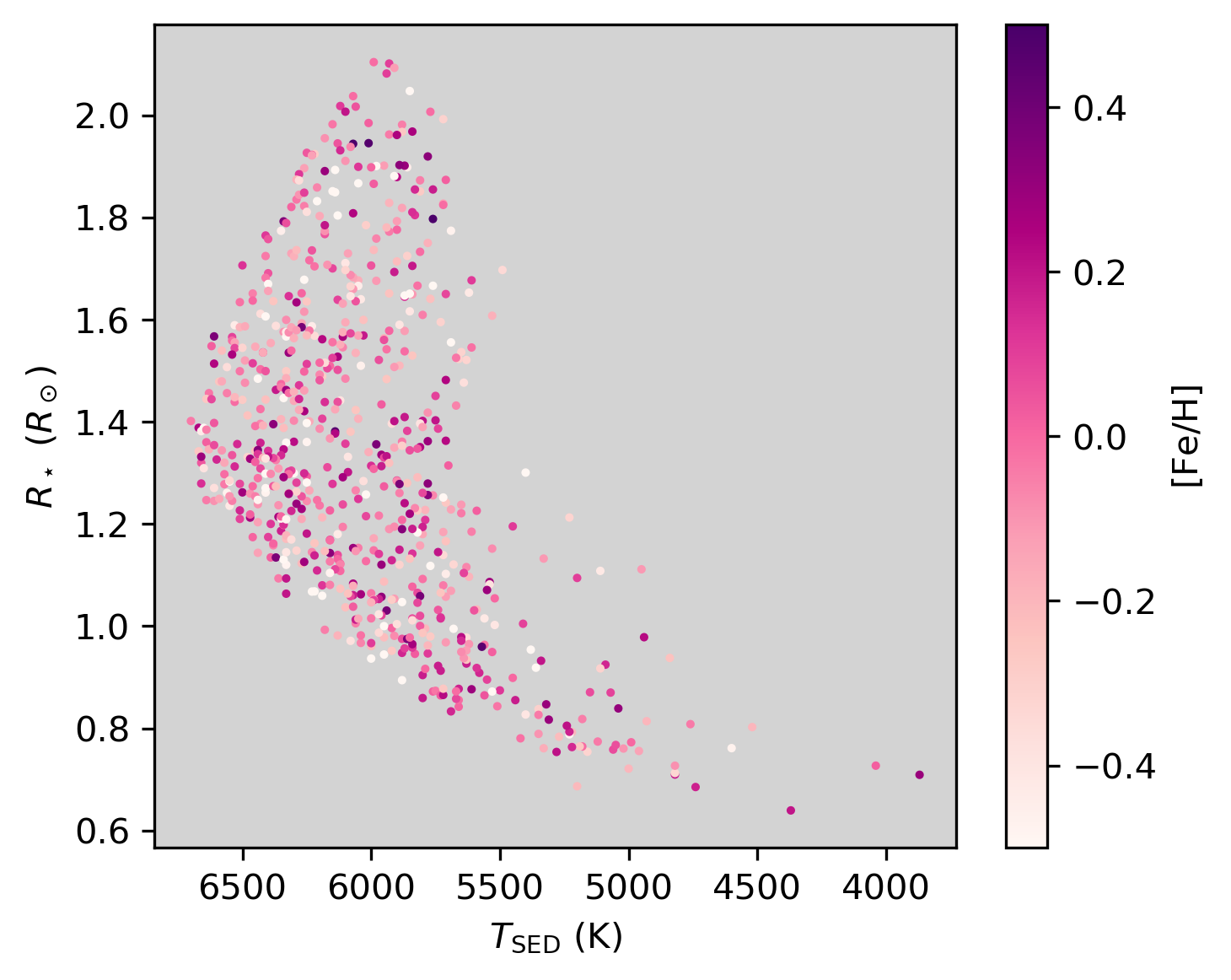}
    \caption{Our sample with the $T_{\rm SED}$ and $R_\star$ derived from SED fitting along with their uncertainties. The colour of each point indicates its spectroscopically determined metallicity. To avoid comparing objects whose observed flux may vary, we omitted objects flagged as double or multiple systems (**), Herbig Ae / Be (Ae*), BY Dra Variables (BYDraV*), Eclipsing Binaries (EclBin), Eruptive Variables (Eruptive*), Red-giant branch stars (RGB*), RS CVn Variables (RSCVnV*), spectroscopic binaries (SB*), variable stars (Variable*), young stellar objects (YSO) or Delta Scuti variables (delSctV*). }
    \label{fig:tsed-rad-plot}
\end{figure}

The resulting $T_{\rm SED}$ and $R_\star$ measured using the SED fitting method are plotted for the gr8stars catalogue in \autoref{fig:tsed-rad-plot}. 
We achieved a median statistical uncertainty of 1.0 per cent and 1.9 per cent in $T_{\rm SED}$ and $R_\star$, respectively. 
However, there will be systematic effects due to using only Solar metallicity atmospheres.
Using results of \citet{morrellUsingStellarSpectral} we find that for $4000\,{\rm K} < T_{SED} < 7000\,K$ we can expect such a systematic to lie below our statistical uncertainties for -0.25<[Fe/H]<0.25.
This metallicity range covers 70\% of our spectroscopic sample, and presumably a similar fraction of the sample which have photometrically determined parameters.

\begin{figure}
    \centering
    \includegraphics[width=\linewidth]{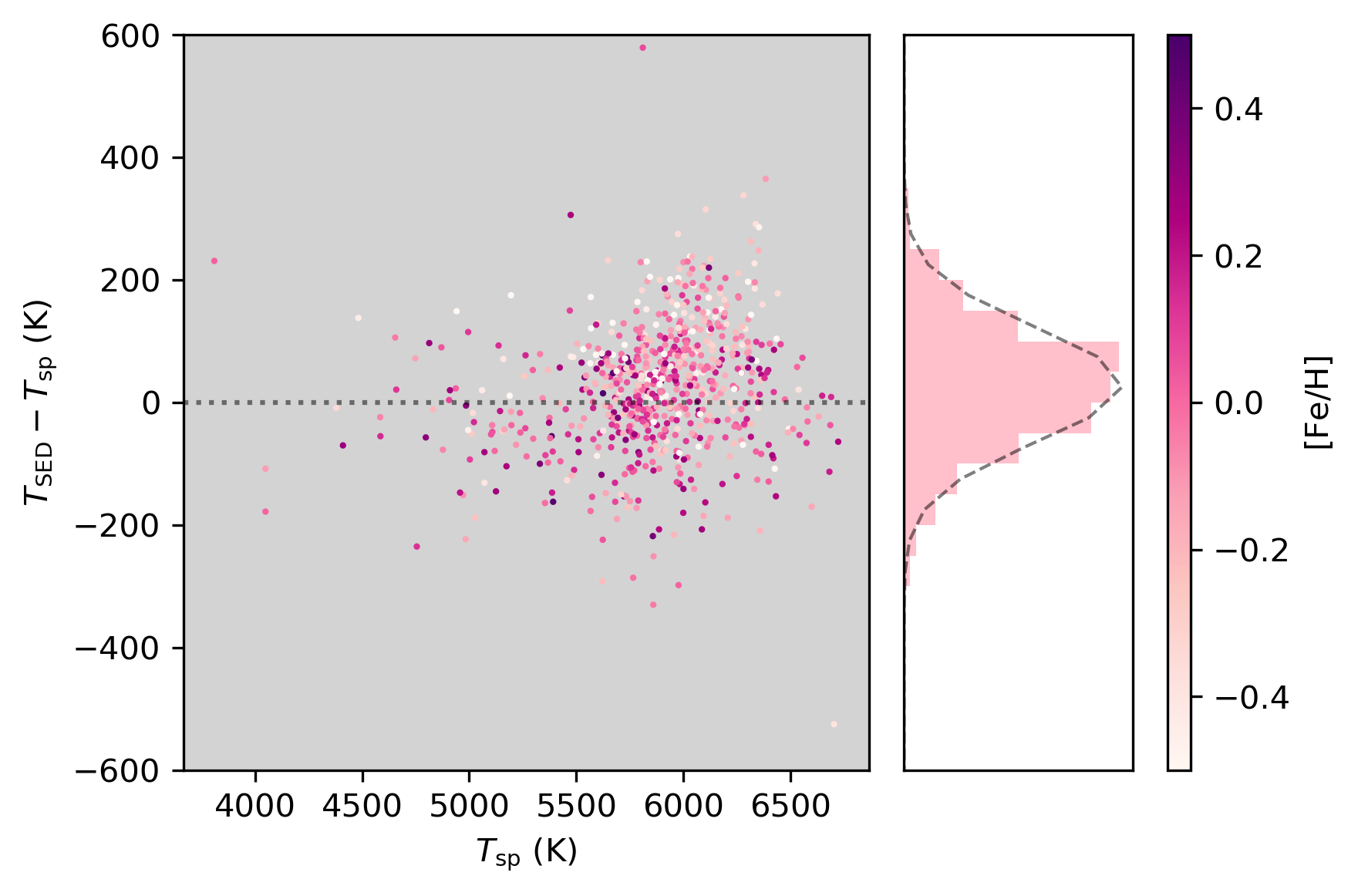}
    \caption{The residuals between the $T_{\rm SED}$ and $T_{\rm sp}$, plotted as a function of $T_{\rm sp}$. The distribution of these residuals, overlaid with a Gaussian fit, are shown in the right hand pane. We omitted from this plot the same targets as specified in \autoref{fig:tsed-rad-plot}. }
    \label{fig:tsed-tsp-comp}
\end{figure}

With the SED-based methods we have achieved a second, orthogonal measurement of temperature. 
A comparison of the temperatures determined from both methods is shown in \autoref{fig:tsed-tsp-comp}. 
The median difference between the temperatures from both methods is 34 K, with the Gaussian fit to the residuals yielding a $\sigma = 98 $ K. 
Although there is some evidence for a systematic with temperature, the scatter is consistent with our cited uncertainties for both methods. 

\subsubsection{Comparison to PASTEL Catalogue}
To further investigate the reliability of our spectroscopic atmospheric parameters, we compared the effective temperature and metallicities determined in this to those from the PASTEL catalogue \citep{soubiran2016}. From the 1716 targets for which we have determined atmospheric parameters, 1638 have effective temperature results in the PASTEL catalogue, and 1171 have metallicty results. As the PASTEL catalogue is a compilation of literature results from various sources and thus not homogeneous, we chose to take the median of these results for use in our comparison, along with a weighted standard deviation to represent uncertainties. In the cases where the PASTEL catalogue does not quote an uncertainty for a measurement, it was omitted from our comparison.

\begin{figure*}
    \centering\includegraphics[width = \textwidth]{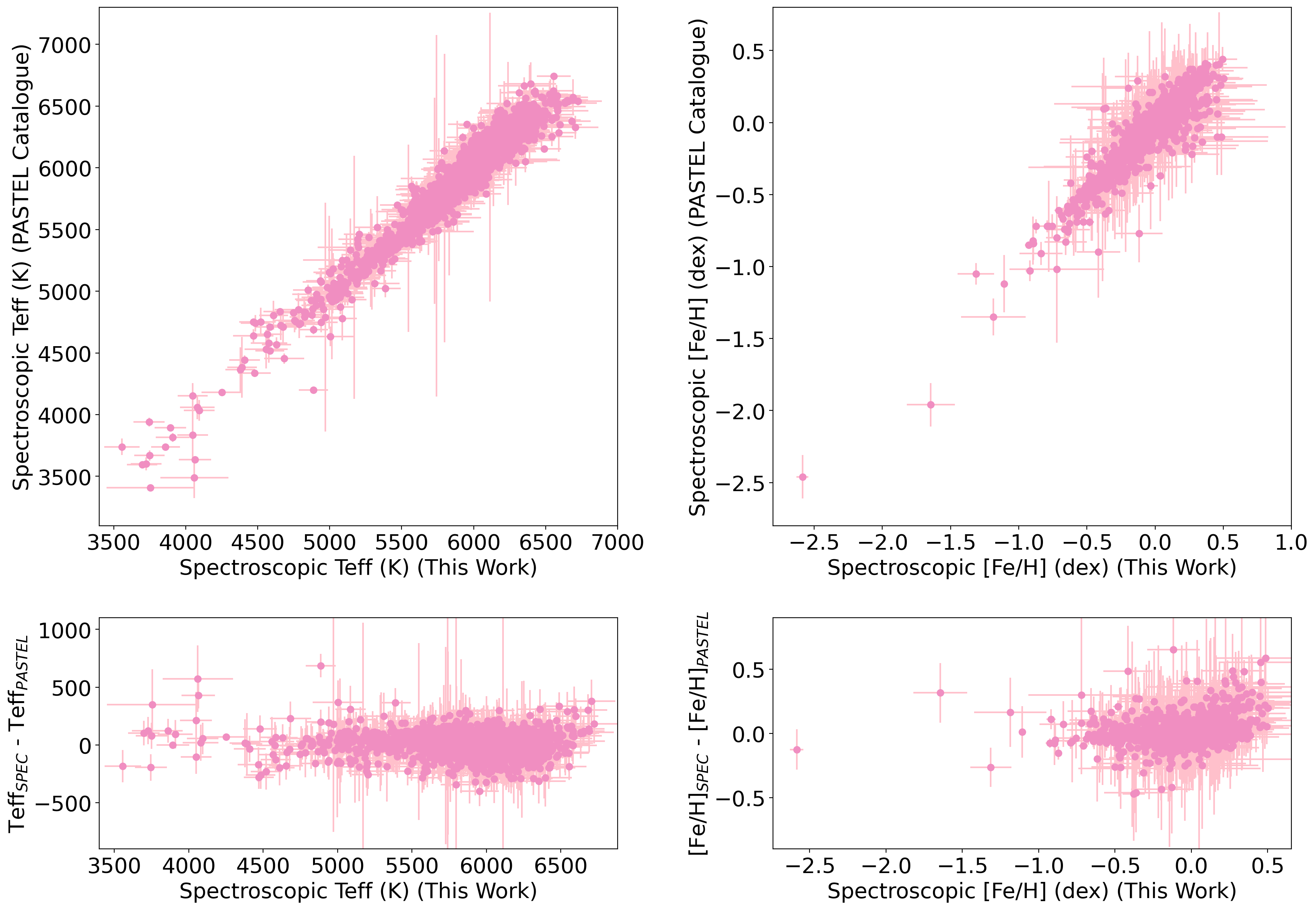}
    \caption{Comparison of spectroscopic effective temperatures (a) and metallicities (b) derived in this work to those from the PASTEL catalogue \citep{soubiran2016}.  The differences between results from this work and those from the PASTEL catalogue are shown in the subplots below.}
    \label{fig:pastel}
\end{figure*}

Figure \ref{fig:pastel} shows the comparison of our results to the effective temperatures and metallicites from the PASTEL catalogue. Overall, the results agree in both parameter spaces. The root-mean-square (RMS) difference between results from this work and those from the PASTEL catalogue are 101 K and 0.11 dex for $T_{\rm eff}$ and $\rm [Fe/H]$, respectively. This matches with our median error of effective temperature and their median error of metallicity. The majority of significant outliers in $T_{\rm eff}$ lie in the cooler regime ($\leq$ 4500 K), with their relatively large uncertainties reflecting the lower SNR of their spectra. HD234677 presents the most prominent outlier in $T_{\rm eff}$ , having a $T_{\rm eff}$ determined in this work to be 4887 $\pm$ 101 K, and 4200 $\pm$ 13 K from PASTEL. The low uncertainty in the PASTEL value suggests high confidence in the result. As the SNR of the spectrum used in this work is relatively low, at SNR = 32, this target will be prioritised in future efforts to complete the \texttt{gr8stars} sample so new spectra can be obtained. 

\subsection{Exoplanet Host Stars}
\begin{figure}
    \centering
    \includegraphics[width=0.5\textwidth]{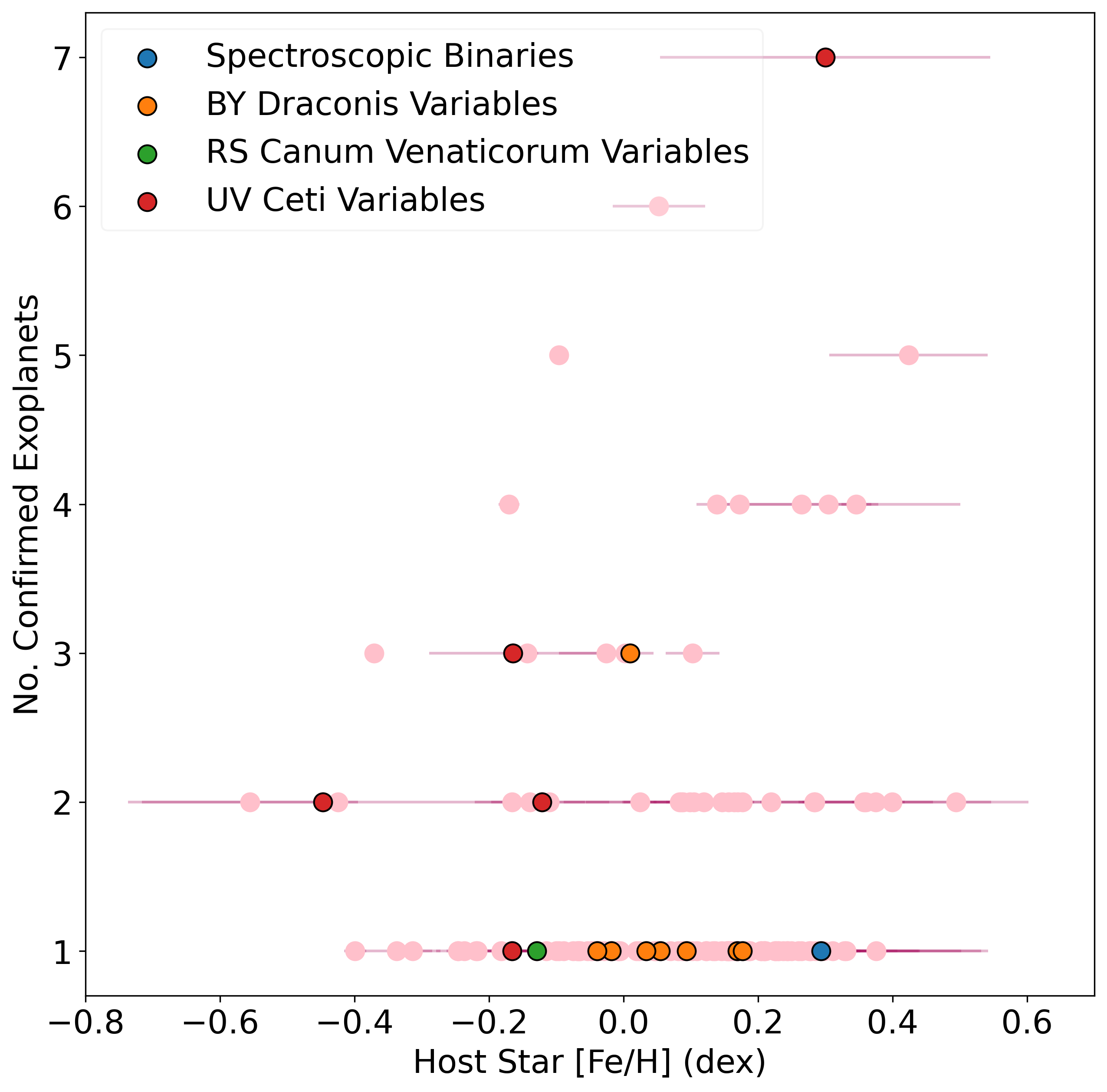}
    \caption{Exoplanet hosts in the \texttt{gr8stars} sample as found from the \texttt{exoplanet.eu} database, plotted as number of confirmed exoplanets vs metallicity of the host star. Special cases are highlighted as in Figure \ref{fig:logg_teff_col}.}
    \label{fig:metal_planets}
\end{figure}

Of the \texttt{gr8stars} targets, 147 were found to be hosts of confirmed exoplanets by cross-matching with the \texttt{exoplanet.eu}\footnote{\url{}https://exoplanet.eu/home/} database. These host stars were found across a range of metallicities, as shown in Figure \ref{fig:metal_planets}. As well-established in the literature, it is expected that higher metallicity stars are more likely to host giant planets than those that are metal-poor \citep[e.g.,][]{santos2004, fischer2005, udry2007, mortier2012, Buchhave2012, buchhave2014}. Evidence of this trend influencing that seen for all planet hosts is apparent in Figure \ref{fig:metal_planets}.

\subsection{Galactic Classification}

\begin{figure}
    \centering
    \includegraphics[width=0.5\textwidth]{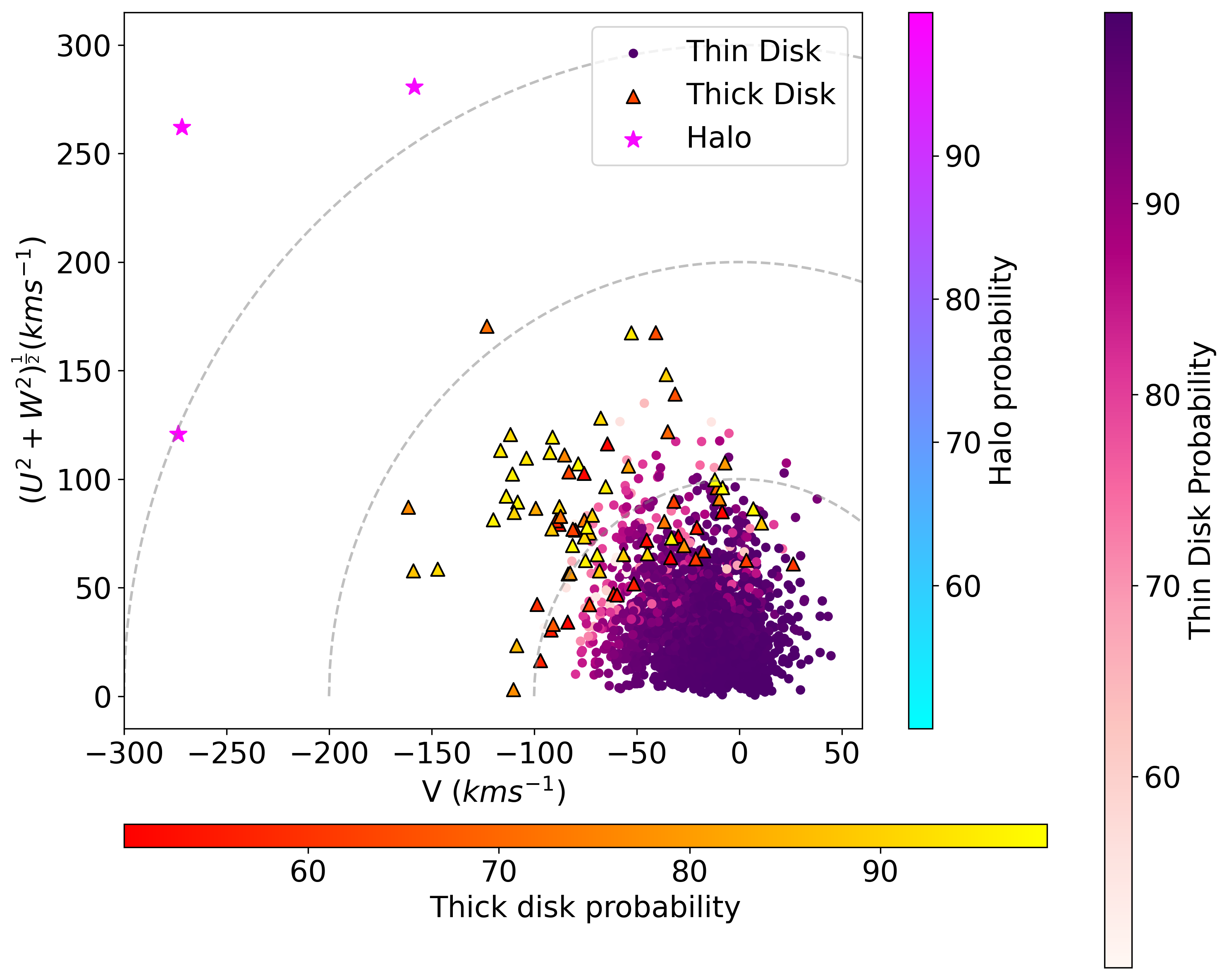}
    \caption{The Toomre diagram of the northern-observable \texttt{gr8stars} sample, with circles representing thin disk stars, triangles representing thick disk stars, and stars representing halo stars. The respective membership groups have individual colour scales representing the probability that each target belongs to that group. Dashed lined show the total velocity ($V_{T}^2 = U^2 + W^2 + V^2$) constants of 100, 200, and 300 $\rm kms^{-1}$.}
    \label{fig:toomre}
\end{figure}

\begin{figure}
    \centering\includegraphics[width = 0.5\textwidth]{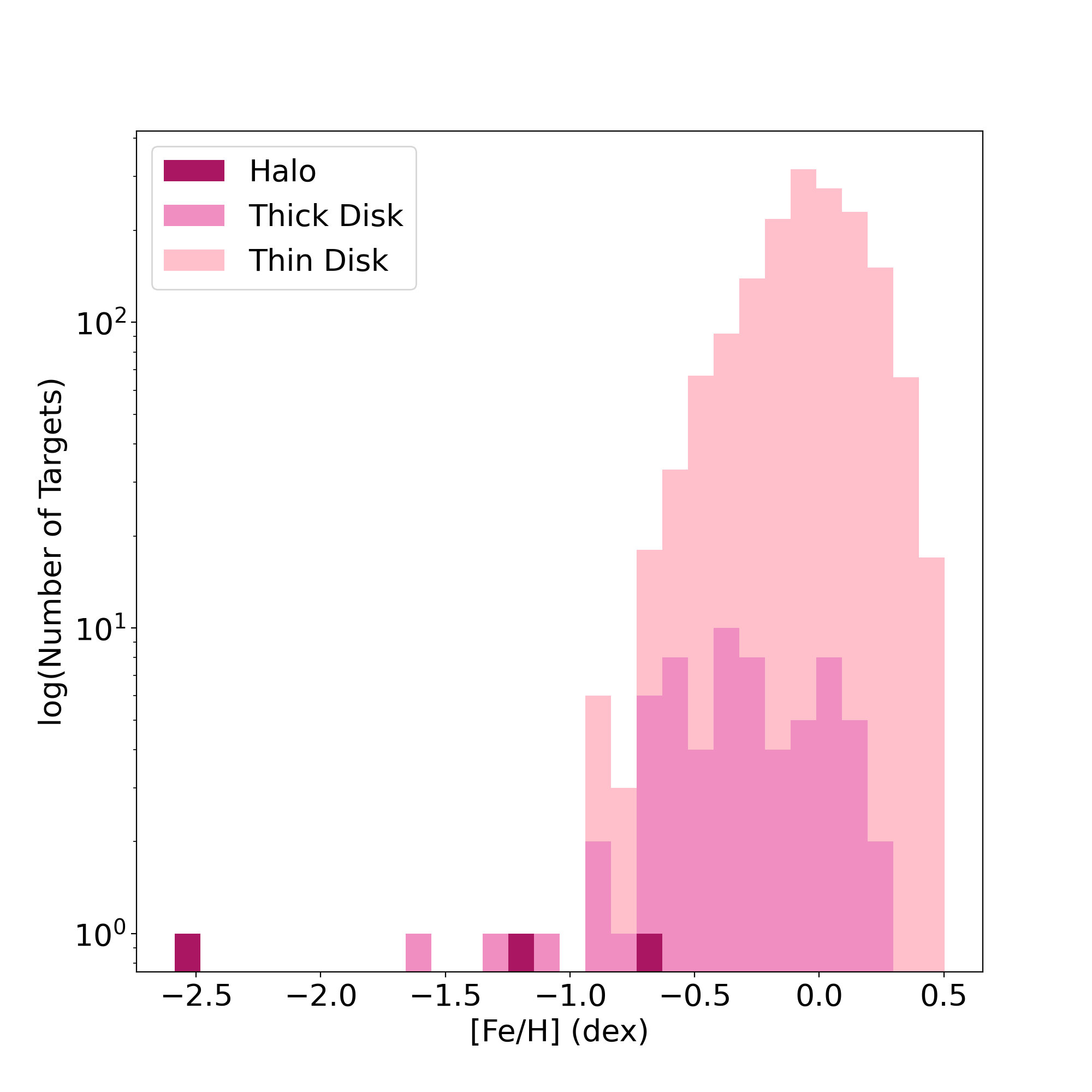}
    \caption{Distribution of metallicities derived from this work spectroscopically in each galactic component group, plotted as a stacked histogram in log scale.}
    \label{fig:mhist}
\end{figure}

Once obtained using the method outlined in Section \ref{galacy_vels}, the galactic velocities were used to determine whether the targets were most likely to be thin disk, thick disk, or halo stars. The most likely membership, along with its probability, for each target was obtained following \citet{reddy2006}. Although all targets are included here, when considering multiple star systems the kinematics may be affected by additional motion \citep{dehnenbinney1998}, therefore reducing the reliability of the membership probability. Within the \texttt{gr8stars} sample, 75.1\% of targets have a Renormalised Unit Weight Error (RUWE) between 0.8 and 1.2, indicating that a single star solution fits well to the astrometric data. Figure \ref{fig:toomre} shows the resulting Toomre diagram of the \texttt{gr8stars} sample, including targets for which we have not yet obtained spectroscopic observations, with the targets marked by their most likely Galactic membership.

Within the entire Northern-observable sample, regardless of whether spectra have been acquired, 2526 targets (96.86\%) are likely to belong to the thin disk, 79 (3.03\%) to the thick disk, and 3 (0.11\%) are likely to be halo stars. A majority thin-disk sample was expected for \texttt{gr8stars}, as all stars are within 103 pc, hence belonging to the solar neighbourhood. Of perhaps the most interest here are the 3 possible halo stars; HD103095\citep{Deliyannis1994, king1997, creevey2012}, HD140283 \citep{bond2013, vandenberg2014, creevey2015, siquiera-mello2015}, and HD148816 \citep{melendez2009}.

The distribution of metallicity in all three galactic component groups is displayed in Figure \ref{fig:mhist} as a stacked histogram. Main-sequence stars in the solar neighbourhood are expected to follow a Gaussian distribution, as shown by \citet{haywood2001}, peaking just below solar metallicity. Observing Figure \ref{fig:mhist}, the distribution of $\rm [Fe/H]$ does indeed follow the expected Gaussian shape for the thin disk stars, centered close to solar metallicity with a mean of -0.04 dex. When considering the thick disk, Figure \ref{fig:mhist} shows the population is far more metal poor than the thin disk, as expected due to the older age of the stars; the mean $\rm [Fe/H]$ for \texttt{gr8stars} thick disk stars is -0.33 dex. We see this again extending to the halo stars; being the oldest stellar components of the galaxy, they are the most metal poor, with a mean $\rm [Fe/H]$ of -1.47 dex.

\section{Conclusions}
\label{conclusion}
The \texttt{gr8stars} catalogue presents a unique database tailored to the purpose of supporting future exoplanet searches in the northern hemisphere. This work collates spectroscopic and photometric parameters for the targets, made publicly available to allow the easy identification and classification of stars. The \texttt{gr8stars} catalogue itself is a library of thousands of spectra for the targets available in both S1D and S2D format, allowing for users to perform their own spectroscopic and radial velocity analyses. 
Observations from multiple instruments are made available in the catalogue for targets where they are present, with their homogeneity in terms of atmospheric parameters derived explored in Figure \ref{fig:inst_comp}. As no systematic trends were present in this testing, the final spectroscopic parameters presented in this work are an inverse-variance-weighted mean for targets with multiple instrument observations. 

While the majority of the Northern-observable targets have available spectra, we will expand the catalogue to include observations for all targets in the sample in the future.

To both explain any outliers in terms of spectroscopic parameters derived, and make users aware of targets deviating from the standard solar-like stars, the special cases identified and first displaying in Figure \ref{fig:logg_teff_col} are noted within the table of final parameters for the catalogue in the `OTYPE' column. The majority of these targets would be less suitable for exoplanet searches than a solar-like star in terms of searching for exoplanets in radial-velocity space, hence it is important that they are identified here. 

Anticipating future work on targets observable from Southern-Hemisphere facilities such as the Second Earth Spectrograph (2ES) under construction for the MPG/ESO 2.2\,m telescope \citep{2024SPIE13096E..8ES}, the target selection process was repeated with the declination constraint replaced with $\delta \le +15\degree$. This produced a list of 2788 targets that will be suitable for such work. Combining this with the previous target list tailored to Northern-hemisphere facilities, and accounting for overlap between the two, the total \texttt{gr8stars} sample encompasses 5645 stars. 

Future papers within the \texttt{gr8stars} collaboration will benefit from a wide range of expertise, covering stellar activity, precision stellar abundances, mass and radius analyses, method comparison and benchmarking, plus many more possible science cases. To facilitate achieving a complete sample for \texttt{gr8stars}, periodic archive searches will be performed for new publicly available spectra. Additionally, current proposals on FIES (PI : Buchhave) and UVES (PI : Freckelton) are currently underway taking observations for both the Northern and Southern samples, and proposals will continue to be submitted as part of the \texttt{gr8stars} collaboration.

Looking to the future, it is vital to consider the importance of \texttt{gr8stars} in the context of revolutionary upcoming missions such as the proposed Habitable Worlds Observatory (HWO) \citep{2021hwo}. Cross-matching our list of 5645 targets with the current proposed star list for HWO \citep{mamajek2024}, we find that 137 of the 164 stars are present in the \texttt{gr8stars} sample. This overlap highlights the significant role that \texttt{gr8stars} could play in the characterisation of host stars for upcoming Earth twin searches.

\section*{Acknowledgements}
We thank the anonymous referee for their constructive report that improved the quality of this paper. 

AVF would like to thank Amaury Triaud for insightful discussions.

This research has made use of data obtained from or tools provided by the portal exoplanet.eu of The Extrasolar Planets Encyclopaedia.
This research has made use of the SIMBAD database, operated at CDS, Strasbourg, France.

Based on observations made with the Nordic Optical Telescope, owned in collaboration by the University of Turku and Aarhus University, and operated jointly by Aarhus University, the University of Turku and the University of Oslo, representing Denmark, Finland and Norway, the University of Iceland and Stockholm University at the Observatorio del Roque de los Muchachos, La Palma, Spain, of the Instituto de Astrofisica de Canarias.

A.M. acknowledges funding from a UKRI Future Leader Fellowship, grant number MR/X033244/1 and a UK Science and Technology Facilities Council (STFC) small grant ST/Y002334/1.

ASM acknowledges financial support from the Spanish Ministry of Science, Innovation and Universities (MICIU) projects PID2020-117493GB-I00 and PID2023-149982NB-I00.

AVF acknowledges the support of the IOP through the Bell Burnell Graduate Scholarship Fund.

BK acknowledges funding from the European Research Council under the European Union’s Horizon 2020 research and innovation programme (grant agreement No 865624, GPRV).

EdM would like to acknowledge support from the UK Science and Technology Facilities Council (STFC, grant number ST/X00094X/1)

GRD acknowledges the support from the UK Space Agency. GRD acknowledges support from the European Research Council (ERC) under the European Union’s Horizon 2020 research and innovation programme (CartographY G.A. n. 804752).

JIGH acknowledges financial support from the Spanish Ministry of Science, Innovation and Universities (MICIU) projects PID2020-117493GB-I00 and PID2023-149982NB-I00.

LLZ gratefully acknowledges support for this work provided by NASA through the NASA Hubble Fellowship grant HST-HF2-51569 awarded by the Space Telescope Science Institute, which is operated by the Association of Universities for Research in Astronomy, Incorporated, under NASA contract NAS5-26555.

M.T. thanks INAF for the support (Large Grant EPOCH), the Mini-Grant PILOT (1.05.23.04.02), and the financial support under the National Recovery and Resilience Plan (NRRP), Mission 4, Component 2, Investment 1.1, Call for tender No. 104 published on 2.2.2022 by the Italian Ministry of University and Research (MUR), funded by the European Union –NextGenerationEU– Project ‘Cosmic POT’ Grant Assignment Decree No. 2022X4TM3H by the Italian Ministry of University and Research (MUR).

NCS acknowledges support by the European Union (ERC, FIERCE, 101052347). Views and opinions expressed are however those of the author(s) only and do not necessarily reflect those of the European Union or the European Research Council. Neither the European Union nor the granting authority can be held responsible for them. This work was supported by FCT - Fundação para a Ciência e a Tecnologia through national funds by these grants: UIDB/04434/2020 DOI: 10.54499/UIDB/04434/2020, UIDP/04434/2020 DOI: 10.54499/UIDP/04434/2020.

S.G.S acknowledges the support from FCT through
Investigador FCT contract nr.CEECIND/00826/2018 and POPH/FSE (EC).the update to pyaneti to do model comparison for different multi-dimensional GPs - is that included in the live version? As we may benefit for our system here.

The Center for Computational Astrophysics at the Flatiron Institute is supported by the Simons Foundation.

\section*{Data Availability}
All stellar parameters determined and uniformly formatted spectra used within this work can be accessed via Zenodo\footnote{\url{https://zenodo.org/records/15441644}}. The table of stellar parameters will additionally be uploaded to Vizier CDS.
Spectra are available upon request to the authors.



\bibliographystyle{mnras}
\bibliography{gr8stars1} 




\appendix

\section{Programme IDs}
\label{progids}
\subsection{FEROS}
082.A-9007, 078.A-9059, 088.C-0892, 0108.A-9029, 0102.A-9008, 074.A-9002, 082.C-0446, 0100.A-9022, 0108.A-9032, 0103.A-0910, 095.A-9029.
\subsection{FIES}
64-302, 55-006, 59-802, 58-901, 61-510, 54-413, 61-605, 59-901, 61-803, 58-199, 63-901, 55-901, 64-201, 55-019, 50-410, 66-010
\subsection{HARPS}
183.C-0972(A), 072.C-0488(E), 099.D-0717(A), 091.C-0034(A), 099.C-0458(A), 0101.C-0379(A), 089.C-0732(A), 0103.C-0432(A), 0100.C-0097(A), 094.D-0596(A), 085.C-0019(A), 093.C-0409(A), 090.C-0421(A), 087.C-0831(A), 097.C-0561(A), 075.C-0332(A), 076.C-0279(A), 080.C-0712(A), 082.C-0412(A), 184.C-0815(F), 080.C-0664(A), 184.C-0815(B), 077.C-0295(A), 081.C-0774(A), 184.C-0815(E), 077.C-0295(B), 076.C-0279(B), 184.C-0815(C), 076.C-0279(C), 198.C-0836(A), 192.C-0852(A), 074.C-0364(A), 188.C-0265(M), 188.C-0265(I), 188.C-0265(R), 188.C-0265(A), 188.C-0265(P), 188.C-0265(G), 188.C-0265(D), 188.C-0265(O), 188.C-0265(L), 188.C-0265(K), 188.C-0265(H), 0100.D-0444(A), 096.D-0402(A), 
096.C-0499(A), 183.D-0729(A), 077.C-0101(A), 188.C-0265(C), 087.C-0012(B), 089.C-0415(B), 087.C-0012(A), 089.C-0415(A), 0101.C-0232(A), 097.C-0090(A), 0103.C-0548(A), 0100.C-0414(B), 0102.C-0338(A), 101.C-0232(B), 60.A-9700(G), 0102.C-0584(A), 091.C-0936(A), 196.C-0042(E), 
190.C-0027(A), 196.C-0042, 196.C-0042(D), 192.C-0224(C), 192.C-0224(B), 
082.C-0212(B), 60.A-9036(A), 191.C-0873, 183.C-0437(A), 191.C-0873(A), 188.C-0265(J), 188.C-0265(N), 188.C-0265(E), 188.C-0265(B), 188.C-0265(F), 075.C-0689(B), 184.C-0815(A), 077.C-0295(C), 102.D-0717(A), 077.C-0364(E), 0102.C-0558(A), 092.C-0721(A), 196.C-1006(A), 188.C-0265(Q), 0103.C-0785, 0101.C-0275, 0104.C-0090(A), 0103.C-0206(A), 0102.C-0338(C), 0100.C-0414(A), 0102.C-0338(B), 080.D-0347(A), 082.C-0390(A), 
185.D-0056(D), 183.D-0729(B), 0102.D-0119(A), 0101.C-0275(A), 0101.C-0232(C), 103.C-0548(A), 083.C-0794(A), 083.C-0794(C), 192.C-0224(H), 083.C-0794(B), 192.C-0224(G), 192.C-0224, 185.D-0056(H), 188.C-0779(A), 185.D-0056(B), 076.C-0155(A), 078.D-0071(C), 079.D-0075(D), 075.D-0194(A), 081.D-0065(D), 076.D-0130(A), 076.D-0130(C), 078.D-0071(B), 072.C-0096(C), 079.D-0075(E), 078.D-0071(A), 072.C-0096(B), 076.D-0130(B), 073.D-0038(D), 079.D-0075(C), 081.D-0065(E), 074.D-0131(C), 078.D-0071(D), 072.C-0096(A), 074.D-0131(A), 079.C-0927(C), 089.C-0497(B), 091.C-0866(C), 098.C-0739(A), 0104.C-0090(B), 288.C-5010(A), 086.C-0284(A), 085.C-0063(A), 078.C-0209(B), 078.C-0209(A), 098.C-0366(A), 192.C-0852(M), 082.C-0212(A), 083.C-0794(D), 0100.C-0474(A), 100.D-0717(A), 0103.D-0445(A), 073.D-0578(A), 075.C-0689(A), 185.D-0056(F), 081.C-0779(A), 086.C-0145(C), 092.C-0427(A), 094.C-0322(A), 
094.C-0322(B), 084.C-0228(A), 082.C-0357(A), 086.C-0145(A), 088.C-0513(B), 075.C-0202(A), 078.C-0044(A), 096.D-0717(A), 095.C-0551(A), 096.C-0460(A), 0100.C-0487(A), 079.D-0075(A), 074.D-0131(E), 081.D-0065(B), 081.D-0065(A), 081.D-0065(C), 072.C-0096(E), 080.D-0086(E), 078.D-0071(E), 079.D-0075(B), 073.D-0038(C), 073.D-0038(B), 072.C-0096(D), 076.D-0130(E), 074.D-0131(B), 080.D-0086(B), 089.C-0739(A), 099.C-0205(A), 098-C-0518(A), 097.C-0571(B), 074.C-0012(A), 0101.C-0106(A), 073.C-0733(C), 185.D-0056(E), 185.D-0056(K), 074.C-0037(A), 078.D-0245(C), 079.C-0127(A), 079.C-0828(A), 081.C-0388(A), 074.D-0131(D), 080.D-0086(C), 080.D-0086(A), 080.D-0086(D), 0103.C-0759(A), 
079.C-0046(A), 077.C-0012(A), 075.D-0760(A), 098.C-0269(A), 098.C-0269(B), 096.C-0876(A), 080.D-0408(A), 0104.C-0418(A), 1101.C-0557(A),183.C-0792, 072.C-0488, 081.C-0148, 089.C-0479, 091.C-0866, 086.D-0082, 106.21DB, 090.C-0395, 088.C-0662, 106.21TJ, 108.22CE, 109.2392, 110.242T, 106.21TJ, 0102.D-0119

\subsection{SOPHIE}
08A.PNP.CONS, 07A.PNP.CONS, 11A.PNP.CONS, 07B.PNP.CONS, 17B.PNPS.AREN, 12B.PNPS.HALB, 16A.PNPS.HALB, 18B.PNPS.HALB, 17B.PNPS.HALB, 17A.PNPS.HALB
% ...

\subsection{UVES}
086.D-0062, 076.B-0055, 072.B-0585, 0104.D-0550, 089.C-0207, 70.D-0356, 094.D-0596, 078.D-0760, 093.A-0373,  089.D-0202, 087.D-0724, 077.B-0507, 108.221X, 087.D-0010, 165.L-0263, 65.L-0507, 109.22VP

\section{FITS header keys}
Table \ref{tab:header} displays the keywords used in the \texttt{gr8stars} FITS file headers.
\begin{table}
    \centering
    \begin{tabular}{cc}
        Header key & Information Contained\\
        \hline
        OBJECT & Primary name of target\\
         INSTRUME & Instrument used to observe the spectrum\\
         RA & Right Ascension of the target\\
         DEC & Declination of the target \\
         EQUINOX & Target equinox\\
         CRPIX1 & Reference Pixel\\
         CRVAL1 & Wavelength at reference pixel \\
         CDELT1 & Wavelength step \\
         GR8 & Date and author of merging \\
         ESTSNR & Estimated SNR of the spectrum \\
    \end{tabular}
    \caption{The keys contained within the headers of the \texttt{gr8stars} S1D spectra.}
    \label{tab:header}
\end{table}

\section{SIMBAD Object Types}
Table \ref{tab:simbads} displays the definitions of the SIMBAD object classifications used throughtout this work.
\begin{table}
    \centering
    \begin{tabular}{c|c}
        OTYPE & Description \\
        \hline
         SB* & Spectroscopic Binary \\
         HighPM* & High Proper Motion Star \\
         BYDraV* & BY Dra Variable \\
         Star & Star \\
         RSCVnV* & RS CVn Variable \\
         ** & Double or Multiple Star \\
         ChemPec* & Chemically Perculiar Star \\
         UV Cet & UV Ceti Variable Star \\
         EclBin & Eclisping Binary \\
         EmLine* & Emission Line Star \\
         RotV* & Rotating Variable \\
         RGB* & Red Giant Branch Star \\
         Variable* & Variable Star \\
         Ae* & Herbig Ae/Be Star \\
         YSO & Young Stellar Object \\
         TTauri* & T Tauri Star \\
         delSctV* & Delta Sct Variable
    \end{tabular}
    \caption{Definition of the SIMBAD object types used to classify the \texttt{gr8stars} targets.}
    \label{tab:simbads}
\end{table}


\bsp	
\label{lastpage}
\end{document}